\documentclass[12pt,preprint,natbib]{aastex}

\RequirePackage{lineno} 
\usepackage{longtable}
\usepackage{epstopdf}  
\usepackage{ifthen}
\usepackage{rotating}
\usepackage{subfigure} 
\usepackage{booktabs}

\def\afrho{{$A$($\theta$)$f\rho$}}
\def\deg{$^\circ$}
\def\rh{$r_{\mathrm{H}}$}

\begin{document}
\bibliographystyle{aj}

\setlength{\footskip}{0pt} 

\title{A Quarter-Century of Observations of Comet 10P/Tempel 2 at Lowell Observatory: Continued Spin-Down, Coma Morphology, Production Rates, and Numerical Modeling}

\author{Matthew M. Knight\altaffilmark{1,2,3}, David G. Schleicher\altaffilmark{2}, Tony L. Farnham\altaffilmark{4}, Edward W. Schwieterman\altaffilmark{1,4}, Samantha R. Christensen\altaffilmark{1}}

\author{Submitted to \textit{The Astronomical Journal} August 5, 2012; Accepted September 14, 2012}

\altaffiltext{1}{Contacting author: knight@lowell.edu.}
\altaffiltext{2}{Lowell Observatory, 1400 W. Mars Hill Rd, Flagstaff, AZ 86001, U.S.A.}
\altaffiltext{3}{The Johns Hopkins University Applied Physics Laboratory, 11100 Johns Hopkins Road, Laurel, Maryland 20723}
\altaffiltext{4}{Department of Astronomy, University of Maryland, College Park, MD 20742-2421}
\altaffiltext{5}{Department of Astronomy, University of Washington, Seattle, WA 98195-1580}

\begin{singlespace}

\newpage
\section*{Abstract}
We report on photometry and imaging of Comet 10P/Tempel 2 obtained at Lowell Observatory from 1983 through 2011. We measured a nucleus rotation period of 8.950 $\pm$ 0.002 hr from 16 nights of imaging acquired between 2010 September and 2011 January. This rotation period is longer than the period we previously measured in 1999, which was itself longer than the period measured in 1988, and demonstrates that Tempel 2 is continuing to spin down, presumably due to torques caused by asymmetric outgassing. A nearly linear jet was observed which varied little during a rotation cycle in both R and CN images acquired during the 1999 and 2010 apparitions. We measured the projected direction of this jet throughout the two apparitions and, under the assumption that the source region of the jet was near the comet's pole, determined a rotational pole direction of RA/Dec = 151{\deg}/$+$59\deg\ from CN measurements and RA/Dec = 173{\deg}/$+$57\deg\ from dust measurements (we estimate a circular uncertainty of 3\deg\ for CN and 4\deg\ for dust). Different combinations of effects likely bias both gas and dust solutions and we elected to average these solutions for a final pole of RA/Dec = 162\deg\ $\pm$ 11{\deg}/$+$58\deg\ $\pm$ 1{\deg}. Photoelectric photometry was acquired on 3 nights in 1983, 2 nights in 1988, 19 nights in 1999/2000, and 10 nights in 2010/2011. The activity exhibited a steep ``turn-on'' $\sim$3 months prior to perihelion (the exact timing of which varies) and a relatively smooth decline after perihelion. The activity during the 1999 and 2010 apparitions was similar; limited data in 1983 and 1988 (along with $IUE$ data from the literature) were systematically higher and the difference cannot be explained entirely by the smaller perihelion distance. We measured a ``typical'' composition, in agreement with previous investigators. Monte Carlo numerical modeling with our pole solution best replicated the observed coma morphology for a source region located near a comet latitude of $+$80$^\circ$ and having a radius of $\sim$10$^\circ$. Our model reproduced the seasonal changes in activity, suggesting that the majority of Tempel 2's activity originates from a small active region located near the pole. We also find that a cosine-squared solar angle function gives the best fit as compared to a standard cosine function.

\begin{description}
\item{\textbf{Keywords}:} comets: general -- comets: individual (10P/Tempel 2) -- methods: data analysis -- methods: observational
\end{description}

\section{INTRODUCTION}
With over two dozen apparitions since its discovery in 1873,
10P/Tempel 2 is one of the better studied Jupiter-family comets. 
Tempel 2 is frequently 
considered as a target for a spacecraft mission (cf. \citealt{osip92}), 
making continued investigations highly desirable.
It has long been known that its total brightness is much higher 
after perihelion as compared to before perihelion (cf. \citealt{sekanina79}), 
and its very anemic outgassing earlier than 2--3 months prior to perihelion 
has made Tempel 2 a subject of nuclear studies in recent decades. 
Its $\sim$5.5 yr period causes it to alternate between favorable
and unfavorable apparitions.
Tempel 2 was particularly well placed for pre-perihelion observations 
in 1988 (and again in 1999), resulting in both rotational period 
measurements \citep{jewitt88,jewitt89,ahearn89,wisniewski90} and values of the albedo, size, and shape \citep{ahearn89}. 
A perturbation by Jupiter in 2002 altered its orbit such that it 
had excellent observational circumstances following perihelion at 
its most recent, 2010, apparition. 

Tempel 2 became only the second comet showing evidence for a change 
in rotation period, following Comet Levy (1990c) \citep{schleicher91,feldman92a}, 
when \citet{mueller96} reported several possible periods 
from 1994 observations -- none of which matched the precise 1988 results. 
Based on their findings, we conducted an observing campaign in 1999 
before significant coma appeared to measure a nuclear lightcurve and 
determine a definitive period. These data were eventually reduced 
and analyzed prior to the 2010 apparition and, indeed, produced a 
unique solution of 8.941 $\pm$ 0.002 hr (\citealt{knight11a}, henceforth 
Paper 1), consistent with one of the five possibilities derived by Mueller 
and Ferrin; their remaining four solutions were ruled out as aliases. 
Our reanalysis of the various 1988 pre-perihelion lightcurves confirmed the 
period solutions of each dataset 
\citep{jewitt88,jewitt89,ahearn89,wisniewski90} and yielded an ensemble 
answer of 8.932 $\pm$ 0.001 hr that was consistent
with those found previously by \citet{sekanina91} and \citet{mueller96}. 
There was therefore a spin down of $\sim$32 seconds 
over an interval of two perihelion passages. If the change was due to 
some type of discrete event (such as a piece of the nucleus breaking 
off) then one would expect no further change (or other changes by 
random amounts if additional events occurred). However, if the spin down 
was due to a systematic torque from ongoing outgassing, then we predicted 
that the period should increase again by $\sim$32 seconds prior to the 2010 
perihelion passage (two additional apparitions) or by $\sim$48 seconds 
following the 2010 apparition. A new observing campaign was conducted in 
2010 to test this hypothesis; the results of 16 nights of lightcurve 
measurements are presented in Section~\ref{sec:rot_period}.

We also performed CCD imaging throughout the 1999 and 2010 apparitions 
to measure coma morphology and to search for variations due to rotation, 
season, and viewing orientation. In addition to the strongly 
asymmetric brightness variations about perihelion, Tempel 2 has 
also been long-known to exhibit a persistent fan or jet. \citet{sekanina79} 
modeled the position angle of this near-linear feature from 
seven apparitions during a half-century to derive a solution for 
the orientation of the rotation axis, and then he derived a significantly 
different solution using different and more appropriate assumptions regarding 
the cause of fan/jet structures based on findings from the Halley spacecraft 
flybys \citep{sekanina87b}. His latter solution was in reasonable agreement 
with his subsequent model answer based on 1988 fan measurements 
\citep{sekanina91}. Our own coma morphology measurements in 1999 and 2010 were 
intended to provide a homogeneous dataset to test and further constrain 
Tempel 2's pole orientation, and the data and results, including a 
model of the jet, are given in Section~\ref{sec:morphology}. 

The pattern of total brightness variations observed in the past 
strongly suggests a seasonal variation induced by the change in 
the sub-solar latitude along with the location of one or more 
source regions on Tempel 2's nucleus. To better quantify the situation 
and test whether or not the nuclear model solutions of the rotation axis 
and the source of the fan/jets would also naturally yield the 
seasonal effects, we obtained extensive photoelectric photometry 
to measure production rates of the gas and dust throughout both 
the 1999 and 2010 apparitions. These data could then be compared to 
a synthetic production rate curve based on our pole solution and 
jet source location, similar to our successful 19P/Borrelly investigation 
\citep{schleicher03b}. Our narrowband photometry, from the 1983, 1988, 
1999, and 2010 apparitions, is also used to determine improved abundance 
ratios and other chemical properties. Data and results are presented 
in Section~\ref{sec:photometry}. 


\section{OBSERVATIONS AND REDUCTIONS}
\subsection{Instrumentation}
Our multiple and somewhat diverse goals for Comet 10P/Tempel 2 as 
described in the Introduction required a diverse set of techniques 
and instrumentation. We continue to use a traditional photoelectric 
photometer to measure production rates and related properties both 
for continuity with earlier data sets and because we achieve improved 
signal-to-noise for these bulk coma measurements as compared to a CCD. 
However, the nuclear lightcurve measurements are better served with 
a CCD, and the coma morphology measurements require a CCD. Conveniently, 
both goals for the CCD could sometimes be obtained concurrently from 
the same data. All observations obtained during the 1999 and 2010 
apparitions used either the Hall 42-in. (1.1-m) or Perkins 72-in (1.8-m) 
located at Lowell Observatory. In all, 19 nights of photometry were 
obtained in 1999/2000 and 10 nights in 2010/2011, while 25 and 27 nights of 
imaging were obtained, respectively. 
Two of the three nights of photometry obtained 
in 1983 also were with the Perkins telescope, while the other was 
with Lowell's 24-in (0.6-m) telescope located at Perth Observatory 
in Western Australia. 
Finally, both nights of photometry in 1988 were obtained with the 
University of Hawaii's 88-in (2.2-m) telescope at Mauna Kea Observatory. 
A listing of nights of observation and the associated observing 
circumstances, including time relative to perihelion ($\Delta$T), 
heliocentric distance ({\rh}), geocentric distance ($\Delta$), and 
phase angle ($\theta$), are presented in Tables~\ref{t:phot_circ} and 
\ref{t:imaging_circ} for the photometry and the imaging, respectively. 
Note that only morphology data for 1999 have been listed since rotational 
data were already listed in Paper 1.



The detectors employed for the photometry were EMI 6256 photomultiplier 
tubes and, except for the night at Perth Observatory when a DC amplifier 
was used, all photometry used pulse counting systems. The narrowband 
filters were from the International Halley Watch (IHW) sets during the 
1980s \citep{ahearn91}, while the two more recent apparitions were with the Hale-Bopp (HB) 
sets \citep{farnham00}. Equivalent HB filters were used, along with a broadband Cousins R 
filter, for the various CCD observations.

The 1999 imaging data were acquired on the Hall 1.1-m telescope with the TI 800 CCD (all June data and July 16) or the SITe 2K CCD (all other 1999 data). On chip 2$\times$2 binning produced images with a pixel scale of 0.71 arcsec for the TI chip and 1.14 arcsec for the SITe chip. 
All 2010 imaging data were acquired on the Hall 1.1-m with the e2v CCD231-84 chip. On chip 2$\times$2 binning gave a pixel scale of 0.74 arcsec; additional 2$\times$2 binning during processing resulted in a final pixel scale of 1.48 arcsec.

\subsection{Photometer Observations and Reductions}
A standard observational set for photometry consisted of the five gas 
filters (OH, NH, CN, C$_3$, and C$_2$) and two (3650 and 4845 \AA) or 
three (3448, 4450, and 5260 \AA) continuum filters with the IHW or HB 
filter set, respectively. Because of our desire to extend the temporal 
coverage as much as possible, however, sometimes a subset of filters was 
used due to the faintness of the comet. Even fewer filters were utilized 
when the primary goal of the photometry was to measure rotational 
variability of the nucleus and/or inner coma, as was the case for the two 
nights in 1988 and the first six nights of 1999. Since the 1988 lightcurve 
data served their purpose and were published \citep{ahearn89}, we 
only include here the few sets that included gas species measurements. 
In the case of the 1999 lightcurve photometry, the data were intended 
to supplement the imaging from other nights, but the plate scale forced 
us to use larger aperture sizes than preferred, yielding significant 
coma contamination and resulting in our not including these data in the 
period determination \citep{knight11a}. Here, we only include a 
few representative data sets from each night's lightcurve, along with 
all of the other sets intended for measurements of production rates. 
All told, this results in four data sets from 1983, nine sets from 1988, 
70 sets from 1999/2000, and 21 sets from 2010/11, with photometer entrance 
apertures ranging from 10 to 204 arcsec and projected aperture radii 
ranging from 2900 km up to 87,100 km. 

As usual, all comet photometry was interspersed with sky measurements 
for each filter and standard stars to measure nightly extinction 
and determine absolute flux calibrations.
Basic methodologies and associated coefficients of reductions to 
fluxes and then to abundances follow the detailed descriptions and 
values given in \citet{ahearn95}. Other aspects, such as the 
decontamination of continuum filters caused by the wings of the 
C$_2$ and C$_3$ bands, have been revised with the introduction of the HB 
filter set and subsequently back-applied to the older IHW filter set 
(cf. \citealt{farnham00,farnham05}). 
The fluorescence efficiencies (or $g$-factors), used to convert gas 
fluxes to the number of molecules, that vary with heliocentric 
velocity are listed in Table 1; some of these have been updated as 
described by \citet{schleicher11}. We also continue to use the 
quantity {\afrho} as a measure of dust production but, 
when appropriate, we now apply an adjustment from the observed phase 
angle ($\theta$) to a phase angle from the Sun of 0$^\circ$ (see 
\citealt{schleicher11}).


\subsection{CCD Observations and Reductions}
\label{sec:ccd_reductions}
As discussed in the Introduction, we had several diverse goals when acquiring data on Tempel 2. Observations by a number of authors in 1988 showed that activity began in earnest 70--90 days prior to perihelion (e.g., \citealt{boehnhardt90}). Therefore, we concentrated on obtaining a nucleus lightcurve early in each apparition, acquiring broadband R-band images almost exclusively. Once activity became more pronounced, we took fewer R-band images and additional HB narrowband images in order to study the coma morphology while still monitoring the lightcurve. Due to somewhat different viewing geometries in 1999 and 2010, the majority of the 1999 data were obtained prior to perihelion, while the majority of the 2010 data were obtained after perihelion. 

The HB narrowband filters are parfocal \citep{farnham00}, but have a slightly different focus than for the broadband R filters. When the primary goal was a lightcurve, the telescope was focused for R-band and consequently slightly out of focus for narrowband images; the reverse was true when the primary goal was coma morphology. Most of the 2010 data were obtained concurrently with observations of 103P/Hartley 2 \citep{knight11b,knight12a}, resulting in less frequent temporal coverage than might have otherwise been expected and, occasionally, narrowband focusing when R-band focusing might have been preferable. Typical exposure times varied depending on the brightness of the comet, with 30 s the shortest exposures (R-band) and 600 s the longest (CN, OH). All images were guided at the comet's rate of motion.

We followed standard techniques to remove the bias and apply a flat field on all science frames (comet and standard stars) using Interactive Data Language (IDL) reduction software. On photometric nights we observed Landolt standard stars \citep{landolt09} when R-band images were obtained and HB narrowband standard stars \citep{farnham00} when narrowband images were obtained. These were used to determine flux calibrations and process the narrowband images into pure gas and pure dust images following our standard photometric procedures (cf. \citealt{farnham00}). Tempel 2 has very little (see Section~\ref{sec:photometry}) contamination from the underlying dust continuum in CN images. After verifying this on a few test cases, we concluded that processed (bias removed and flat fielded) but not decontaminated CN images were sufficient to use for the coma morphology studies described in Section~\ref{sec:morphology}. This allowed us to use many more nights of data than would otherwise be possible if we restricted the CN morphology studies to only photometric nights. Owing to the lower contrast relative to the dust for C$_2$ and C$_3$ and a lack of signal for OH, their discussion is restricted to photometric nights. 

Centroiding was performed by fitting a two-dimensional Gaussian to the apparent photocenter of each image. This worked well on the images dominated by dust. However, it produced systematic offsets in the CN images due to a strong radial CN feature (which will be discussed further in Section~\ref{sec:morphology}) and a relatively weak central condensation. This resulted in an apparent CN centroid which was typically offset along the radial feature by 1--3 pixels (as verified by tracking the centroid, corrected for differential refraction, during long stretches on the comet in multiple filters). Thus, we frequently specified the CN centroid manually using a by-eye estimate of the position of the photocenter. This likely resulted in slightly larger uncertainty in the CN centroid, but we estimate this to be less than 1 pixel, and we were mindful of this uncertainty when analyzing features close to the photocenter.

\section{ROTATION PERIOD IN 2010}
\label{sec:rot_period}

\subsection{Photometric Extractions and Adjustments}
In order to measure the rotation period, we conducted aperture photometry on all broadband R images to obtain the nucleus flux. The general method is similar to that utilized in Paper 1 for the lightcurve data in 1999. We centroided on the nucleus using a two-dimensional Gaussian fit then extracted the flux in a series of circular apertures from 1,2,3,...10 pixels (1.47--14.7 arcsec) in radius. The sky background was determined in an annulus with inner and outer radii of $\sim$51 and $\sim$66 arcsec, respectively. We used the 5.9 arcsec radius aperture for photometric analysis as it gave the most coherent lightcurve, including nearly all of the light even when the seeing was poor or the nucleus PSF was large. This also approximately matched the aperture we used on the 1999 data in Paper 1 (6 arcsec). On non-photometric nights, we applied the extinction corrections and absolute calibrations from neighboring photometric nights in order to determine an approximate absolute calibration. The deviation of the brightness from photometric nights helps give an estimate of how much obscuration was affecting the non-photometric nights. Furthermore, we are primarily interested in the shape of the lightcurve, so systematic offsets from one night to the next are relatively unimportant. 

We measured the brightness of 10 field stars on each night to monitor the varying obscuration during the night. Selected comparison stars remained in the field of view throughout the night while not passing close enough to the comet to have significant coma contamination. We avoided stars which saturated or were very faint, with nearly all comparison stars between magnitude 11 and 16. We used these comparison stars to correct the comet's magnitude during each night as detailed in Section 2.3 of Paper 1. Typical comparison star corrections during a night were less than 0.02 mag on nights deemed to be photometric or to have thin cirrus (see Table~\ref{t:imaging_circ}), and ranged from 0.05 mag to more than 1.0 mag for nights noted as having clouds. We excluded all images with comparison star corrections in excess of 0.3 mag from the lightcurve analysis.

Since our 2010 lightcurve data were all obtained post-perihelion, they suffered from more coma contamination than did our 1999 data. We estimated the coma in the photometric aperture in a similar manner as in Paper 1, using the total annular flux for apertures from 4, 5, 6,... 10 pixels in radius to estimate the flux inside of a radius of 4 pixels (5.9 arcsec). We then removed the average estimated coma for all images during a night (excluding those with obvious contamination from stars and/or cosmic rays). This assumes that coma variations during a night are minimal. The morphology analysis (discussed in Section~\ref{sec:morphology}) shows that the primary jet activity is at the pole, and thus the coma variations due to that jet during a rotation are minimal. Any coma variations produced by a smaller, undetected jets are likely to be inconsequential and thus should not introduce any significant shifts in the timing of the nucleus lightcurve.
The removed coma typically represented 60--80\% of the flux in the photometric aperture, with a higher fraction early in the apparition (since there was more coma closer to perihelion). Removing the coma significantly increased the lightcurve amplitudes, from 0.11--0.24 mag before coma removal to 0.48--0.80 mag after coma removal.

Next, we applied the standard asteroid normalization to correct for changing viewing circumstances, 
\begin{equation}
\label{e:mag_norm}
m_{R}(1,1,0) = m_{R\mathrm{,cr}} - 5\mathrm{log}(r_\mathrm{H}\Delta) - \beta \theta,
\end{equation}
where $m$$_R$(1,1,0) is the normalized magnitude at $r_\mathrm{H}$ = $\Delta$ = 1 AU and $\theta$ = 0$^\circ$, $m_{R\mathrm{,cr}}$ is the apparent magnitude, $m_R$ (which has had the absolute calibrations, extinction corrections, and comparison star corrections applied), with the coma removed, $r_\mathrm{H}$ is the heliocentric distance (in AU), $\Delta$ is the geocentric distance (in AU), $\beta$ is the linear phase coefficient ($\beta$ = 0.032 mag deg$^{-1}$ to match that used in Paper 1), and $\theta$ is the phase angle. Equation~\ref{e:mag_norm} removes the secular variation in brightness and allows comparison of all lightcurves with similar geometries on a similar scale. The geometric corrections are given as $\Delta$$m_1$ in column 10 in Table~\ref{t:imaging_circ} at the midpoint of each night's observations. Finally, we adjusted each night's lightcurve by $\Delta$$m_2$ (given in column 11 in Table~\ref{t:imaging_circ}) in order to make each run's lightcurve peak at the same brightness. Slight variations in peak brightness from night to night are due to the application of absolute calibrations on non-photometric nights, comparison stars that are normalized to their brightest point in the night rather than a catalog value, and the shape of the coma removed. The adjustment is an arbitrary adjustment to aid in lightcurve comparison and should not be interpreted as any particular physical property. As discussed below, we use the timing of the maxima and minima to determine a rotation period, so slight adjustments in the magnitude of all points on a night have no effect on the measured rotation period.

We provide the final photometry in Table~\ref{t:photometry}. Columns 1 and 2 are the UT date and time (at the telescope) at the midpoint of each exposure. Column 3 is $m_R$, the observed R-band magnitude after photometric calibrations, extinction corrections, and comparison star corrections have been applied. Column 4 is $m_R$$^*$, the coma-removed, reduced magnitude $m_R$(1,1,0) corrected by $\Delta$$m_2$ so that all nights have a similar peak magnitude. We obtained 784 usable data points after removing points due to contamination from background stars, tracking problems, cosmic ray hits, or excessively high extinction. While lightcurve data were obtained from 2010 March to 2010 August and on 2011 January 7 and 9, they were of insufficient quality and/or too short duration to be useful in determining a lightcurve and were excluded. Typical scatter in the data used for lightcurve analysis was 0.02--0.03 mag for the data from September through November and 0.05--0.10 in December and January (when the comet was significantly fainter).


\subsection{Measured Rotation Period}
\begin{figure}
  \centering
  \plotone{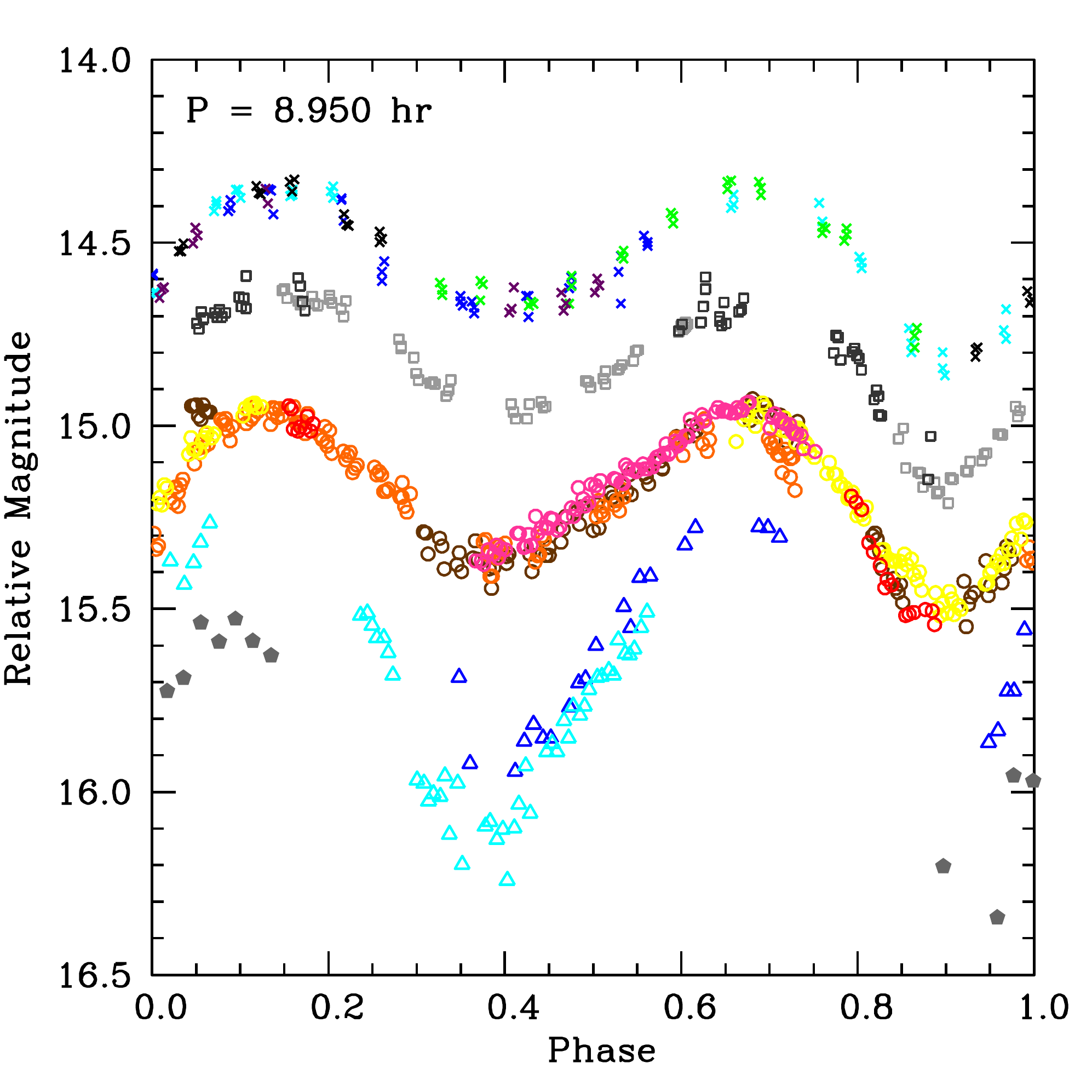}
  \caption[2010 data phased to best solution]{2010 lightcurve data phased to 8.950 hr. The data are the $m_R^*$ data given in column 4 of Table~\ref{t:photometry} (normalized to  {\rh} = $\Delta$ = 1 AU, phase angle = 0{\deg}, coma contamination removed, and offset by $\Delta$$m_2$ so that all nights peak at the same brightness) with additional offsets applied to shift later lightcurves lower in the figure so they do not overlap. The October data were shifted down by 0.3 mag, November by 0.6 mag, December by 0.9 mag, and January by 1.2 mag.
Crosses are September 9 (blue), September 10 (cyan), September 11 (green), September 12 (purple), September 13 (black); squares are October 16 (light gray), October 17 (dark gray); circles are November 2 (brown), November 3 (orange), November 4 (yellow), November 5 (pink), November 7 (red); triangles are December 9 (blue), December 10 (cyan); the filled gray pentagons are January 11. Zero phase was set to perihelion.}
  \label{fig:phased_good}
\end{figure}

\begin{figure}
  \centering
  \plotone{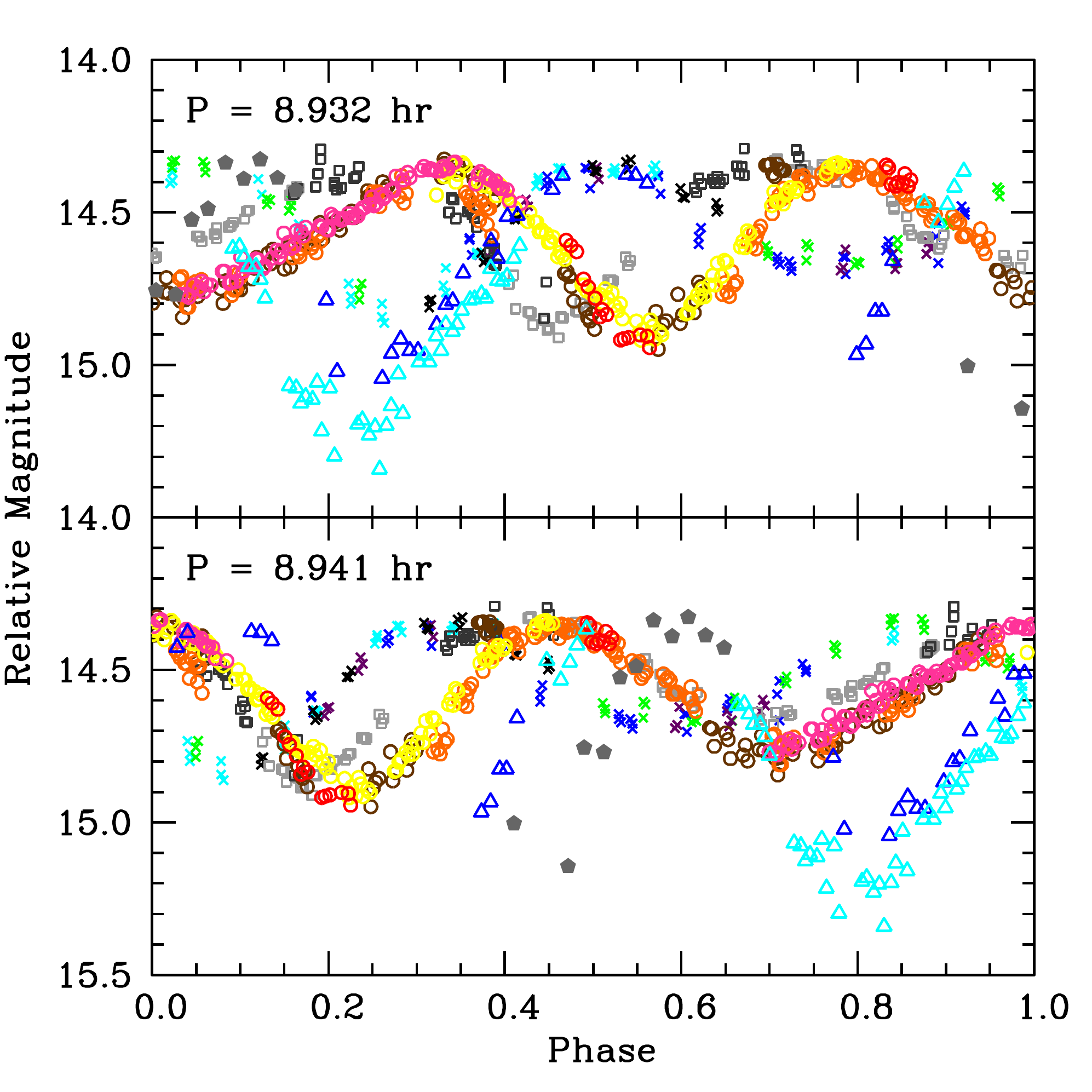}
  \caption[2010 data phased to 1988 and 1999 solutions]{2010 lightcurve data phased to 8.932 hr (the 1988 period, top) and 8.941 hr (the 1999 period, bottom). These solutions (from \citealt{knight11a}) are clearly incompatible with the 2010 data. Symbols are as given in Figure~\ref{fig:phased_good}. The magnitudes are the $m_R^*$ as given in column 4 of Table~\ref{t:photometry}.}
  \label{fig:phased_bad}
\end{figure}

In order to phase the lightcurve data to determine the rotation period, we defined zero phase to be perihelion (2010 July 4.907) and accounted for the light travel time (Column 7 of Table~\ref{t:imaging_circ}). As described further in Paper 1, we used an interactive period search routine which updates the lightcurves on the fly to scan through potential periods. Using the alignment of the peaks and troughs to gauge the goodness of a solution, we found a double-peaked solution (shown previously to be the only viable shape) of 8.950 $\pm$ 0.002 hr. We estimated the uncertainty to be the smallest offset at which the lightcurves were visibly out of phase. The phased lightcurve is plotted in Figure~\ref{fig:phased_good} using $m_R^*$ with runs offset vertically by 0.3 mag from each other so that they do not overlap and can easily be distinguished. The 2010 data were clearly incompatible with either the 1999 period (8.941 $\pm$ 0.002 hr) or the 1988 period (8.932 $\pm$ 0.001 hr), as shown in Figure~\ref{fig:phased_bad} where $m_R^*$ is now shown without offsets so that the whole dataset peaks at the same magnitude. We could not detect a trend in the rotation period in subsets of the data.






\subsection{Sidereal Versus Synodic Rotation Period}
The rotation periods we discussed above are the synodic rotation periods, which do not account for the changing geometry as the Earth and comet move between observations. In Paper 1, we showed that for pole solutions within 20$^\circ$ of that given by \citet{sekanina87b} the difference between the synodic and sidereal periods was smaller than the difference between the synodic periods we measured in 1988 and 1999, thus concluding that the sidereal rotation period had indeed changed during the intervening years. As will be discussed in Section~\ref{sec:morphology}, we used the coma morphology from both 1999 and 2010 to determine a pole only 8.8{\deg} from the pole we assumed in Paper 1. With our pole solution, we then determined the prograde and retrograde sidereal rotation periods necessary to produce the synodic periods we measured from data collected in 1988 \citep{jewitt89,ahearn89,wisniewski90}, 1994 \citep{mueller96}, 1999/2000, and 2010/2011. These are given in Table~\ref{t:sidereal_period}. The 2010/2011 synodic period of 8.950 hr yields a sidereal period of 8.947 hr for the prograde case and 8.953 hr for the retrograde case. The offsets between synodic period and the respective sidereal periods are in the same direction as in 1988, 1994, and 1999. However, the values of the offsets vary somewhat between apparitions because the comet was observed at different positions in the orbit during each apparition. Both the prograde and retrograde solutions confirm that the sidereal period changed between 1988 and 1999 and again between 1999 and 2010. This will be discussed further in Section~\ref{sec:rot_disc}. 


\subsection{Lightcurve Shape and Amplitude}
\label{sec:amplitude}
The lightcurve shape is not symmetric, with the minimum near 0.85 phase clearly deeper and steeper than the minimum near 0.35 phase for data obtained on the same observing run. The same phenomenon was observed in our 1999 data (Paper 1) and allows us to unequivocally identify the same rotational phase in widely separated data. The coma removed lightcurve had peak-to-trough amplitudes of 0.48 mag in 2010 September, 0.50 mag in 2010 October, 0.57 mag in 2010 November, 0.85 mag in 2010 December, and $>$0.8 mag in 2011 January (albeit with large uncertainty given the low number of measurements and large scatter). These magnitudes are similar to those we measured in Paper 1 (0.45--0.75 mag) and those reported by previous authors from 1988 data (0.5--0.8 mag; \citealt{ahearn89}; \citealt{jewitt89}; \citealt{wisniewski90}). However, the viewing geometry was different in 2010 than in 1988 or in 1999, and our amplitudes in November, December, and January actually exceed the expected amplitude based on our pole solution (see Section~\ref{sec:morphology}), the \citet{ahearn89} thermal-IR amplitude in 1988 June, and the assumption that the nucleus is a prolate spheroid with axial ratio A:B:C = 2.13:1:1 (with the axial ratio inferred from the A'Hearn et al. amplitude). Since the A'Hearn et al. data were insensitive to the dust in the coma they give the best measure of the true nucleus shape, and suggest that our coma removal technique overestimated the amount of coma contamination. Since we are primarily interested in the shape of the lightcurve in order to properly phase the data (in fact, the same period was measured prior to coma removal), the chosen coma removal is sufficient.

\subsection{Discussion}
\label{sec:rot_disc}
The spin down of Tempel 2 between 1999 and 2010 supports our preferred hypothesis from Paper 1 for the cause of the spin-down between 1988 and 1999, that the changing rotation period is due to ongoing torquing from outgassing rather than a one-time event such as fragmentation or an impact. While a change of rotation period has been observed for a small handful of comets (e.g., 1990c Levy by \citealt{schleicher91} and \citealt{feldman92a}; 6P/d'Arrest by \citealt{gutierrez03b}; 2P/Encke by \citealt{fernandez05}; C/2001 K5 LINEAR by \citealt{drahus06}; 103P/Hartley 2 by, e.g., \citealt{ahearn11b}, \citealt{drahus11}, \citealt{knight11b}, \citealt{meech11}, and \citealt{samarasinha11}), the only other comet for which a change in rotation period has been measured on multiple orbits is 9P/Tempel 1 \citep{belton11}.
The change in period is fairly consistent from orbit to orbit, suggesting that the torques, and presumably the activity, are somewhat repeatable. This should allow accurate predictions for future apparitions, a highly desirable scenario if Tempel 2 is to become a future spacecraft target.

Changes in comet rotation period caused by outgassing have long been advocated on theoretical grounds. \citet{whipple50} argued for spin-up due to torques caused by outgassing while \citet{wallis84} argued for spin-down due to the loss of angular momentum. More recent work (see, e.g., the review by \citealt{samarasinha04}) has shown that outgassing can not only alter the rotation period but also cause non-principal axis rotation. To test that the period change of Tempel 2 is plausibly due to outgassing, we compute the fractional change in the spin angular velocity resulting from mass loss given by \citet{jewitt97}:
\begin{equation}
\frac{\Delta\omega}{\omega} = k_{T}\left(\frac{{\Delta}M}{M}\right)\left(\frac{V_\mathrm{th}}{V_\mathrm{eq}}\right).
\label{eq:spin}
\end{equation}
Here $k_T$ is the dimensionless moment arm (estimated to be 0.05 by \citealt{jewitt97}), $\Delta$$M$/$M$ is the fractional mass loss rate, $V_\mathrm{th}$ is the outgassing velocity, $V_\mathrm{eq}$ = $r_{n}\omega$ is the equatorial velocity, $r_n$ is the effective nucleus radius, $\omega$ = 2$\pi$/$P$ is the angular spin rate, and $P$ is the rotation period. We assume $M$ = (4$\pi$/3)$r_n^3\rho$ is the mass of the nucleus for a bulk density $\rho$ and estimate the mass lost, $\Delta$$M$, by integrating the H$_2$O production rate curve. 
We use $r_n$ = 5.98 km \citep{lamy09}, 
$\Delta$$M$ = 4.0$\times$10$^{12}$ g (see Section~\ref{sec:photometry}), $P$ = 8.95 hr, $\Delta$$P$ = 0.0034 hr (the average change in period per orbit since 1988), $V_\mathrm{th}$ = 0.7 km s$^{-1}$ (the outgassing rate we estimated from the numerical modeling in Section~\ref{sec:modeling}), and assume $\rho$ = 0.4 g/cm$^3$. This yields $\Delta\omega$/$\omega$ = 0.00033, which compares favorably with the observed $\Delta$$P$/$P$ = 0.00038 and confirms that the change in rotation period is plausibly caused by torquing due to outgassing.

For comparison, we made the same calculation for 9P/Tempel 1 and 103P/Hartley 2 since, as objects of spacecraft flybys, their nucleus properties are much better known. For Tempel 1 we use $r_n$ = 3.0 km \citep{ahearn05a}, $\rho$ = 0.45 g/cm$^3$ \citep{davidsson07,richardson07}, $\Delta$$M$ = 2.9$\times$10$^{12}$ g \citep{schleicher07a}, $P$ $\sim$ 41 hr, $\Delta$$P$ $\sim$ 15 m \citep{belton11}, and $V_\mathrm{th}$ = 1.0 km s$^{-1}$, yielding $\Delta\omega$/$\omega$ = 0.022 and $\Delta$$P$/$P$ = 0.0061. For Hartley 2, we use $r_n$ = 0.58 km \citep{ahearn11a}, $\rho$ = 0.3 g/cm$^3$ \citep{thomas12}, $\Delta$$M$ = 1.5$\times$10$^{12}$ g (here we estimate the mass lost only from mid-August through mid-November, the interval during which the period was measured to change; \citealt{knight12a}), $P$ = 16.5 hr, $\Delta$$P$ = 2 hr \citep{knight11b},\footnote{Note that Hartley 2 is in non-principal axis rotation. We use here the rotation period which apparently dominates the coma morphology and changed most rapidly during the 2010 apparition. An additional component period near 55 hr also exists and may be changing slowly (cf. \citealt{ahearn11b}). If this period and its small change are instead used, $\Delta$$P$/$P$ would be considerably smaller.} and $V_\mathrm{th}$ = 1.0 km s$^{-1}$, finding $\Delta\omega$/$\omega$ = 5.0 and $\Delta$$P$/$P$ = 0.12. This significantly higher $\Delta\omega$/$\omega$ is consistent with the observation that most of Hartley 2's water production comes from the breakup of slow moving icy grains in the coma rather than leaving the nucleus directly at high velocity (cf. \citealt{ahearn11b}). Since Equation~\ref{eq:spin} assumes the mass is leaving the nucleus at $V_\mathrm{th}$ while most of Hartley's water actually leaves the nucleus at a much lower velocity, the majority of the mass lost has a negligible effect on the spin, and $\Delta\omega$/$\omega$ is likely an order of magnitude smaller. 
Despite $\Delta$$P$s that vary by orders of magnitude and very different nuclear sizes, shapes, and outgassing rates, the change in rotation periods of Tempel 2, Tempel 1, and Hartley 2 can all be plausibly explained by the same mechanism: torques due to outgassing. 

The sidereal period (Table~\ref{t:sidereal_period}) increased between the 1988 and 1999 apparitions by 0.008 hr for the prograde solution and 0.010 hr for the retrograde solution, and increased by 0.009 hr between the 1999 and 2010 apparitions for both solutions. The astute reader will notice that while the differences in sidereal periods were similar,
the 1988 and 1999 periods were measured prior to perihelion, whereas the 2010 period was measured after perihelion. Thus, there were two intervening perihelia from 1988 to 1999, but three between 1999 and 2010, and the average change in sidereal period per perihelia is smaller between 1999 and 2010 (0.003 hr/orbit) than between 1988 and 1999 (0.004 hr/orbit in the prograde case and 0.005 hr/orbit in the retrograde case). We note several possible explanations for this apparent discrepancy and a combination of these effects is likely at work. First, the 2010 data were acquired relatively soon after perihelion. As will be shown in Section~\ref{sec:photometry}, Tempel 2's production rates peak after perihelion, and the activity turns off more gradually post-perihelion than it turns-on pre-perihelion. Since the spin-down is presumably caused by asymmetric outgassing and more outgassing occurs post-perihelion, it may not have yet achieved the full change in rotation period during the 2010 apparition when our data were acquired. Next, the change in period may have decreased because the outgassing decreased by $\sim$60\% from 1988 to 1999 and 2010 (discussed in Section~\ref{sec:photometry}). Since $\Delta$$P$ is proportional to the outgassing rate, this would naturally explain a smaller $\Delta$$P$ from 1999 to 2010 than was observed from 1988 to 1999. The final explanation is that the uncertainties in the measured synodic periods (and thus in the inferred sidereal periods) each orbit are large enough that the spin-downs are in fact consistent. For example, in the prograde case, using 8.937 hr for the sidereal period in 1999 (within the uncertainties) and 8.948 hr for the 2010 period (also within our uncertainty) yields an average spin-down of $\sim$0.0035 hr/orbit during both intervals. 

While the sidereal period for the retrograde solution increased by 0.005 hr/orbit between 1988 and 1999, it increased by 0.003 hr/orbit when considering the 1988 and 1994 data (the 1994 data were acquired post-perihelion so there were two intervening perihelion passages since the 1988 data).
This is identical to the spin-down from 1999 to 2010 and, at first glance, appears to avoid the discrepancy discussed in the preceding paragraph. However, there is a difference in sidereal periods of 0.004 hr between the 1994 and 1999 data with no intervening perihelion passage between these data. This is larger than the inferred change from one orbit to the next and is therefore considered unlikely since torquing is assumed to be strongest near perihelion. Thus, the retrograde sidereal periods can be reconciled with each other only if one or more of the measurements are near the extremes of their estimated uncertainties. 

Additional measurements of the rotation period in the next few years (while Tempel 2 is still far from perihelion and has not yet had its period affected by activity during the next perihelion passage) could clarify which scenario is correct. If the comet continued to spin-down post-perihelion, we predict the sidereal period would be 8.950 hr $\pm$ 0.004 hr in the prograde case and 8.959 hr $\pm$ 0.004 hr in the retrograde case. Otherwise the sidereal period should be essentially unchanged, or 8.947 $\pm$ 0.002 hr in the prograde case and 8.953 $\pm$ 0.002 hr in the retrograde case. In any event, the sidereal and synodic periods should be virtually identical at this time, since the effects of the changing viewing geometry are minimal near aphelion (see, e.g., the discussion by \citealt{mueller96}).

We have listed the sidereal period for both the prograde and retrograde pole solutions in Table~\ref{t:sidereal_period} and thus far have assumed that either solution is plausible. We see hints that the prograde solution is the correct choice, but the evidence is not yet convincing enough to rule out the retrograde solution. First, the sidereal periods inferred for the 1994 data of \citet{mueller96} and our own 1999 data are equal (8.938 hr) for the prograde solution but different by 0.004 hr (8.940 versus 8.944 hr) for the retrograde solution. Since there were no intervening perihelion passages between these datasets, we would expect them to have nearly identical sidereal periods, thus favoring the prograde solution. Furthermore, for the retrograde case, the difference in sidereal periods between 1988 and 1994 (when there were two intervening perihelion passages) is 0.006 hr, only slightly larger than the difference between 1994 and 1999, again making the retrograde case less likely. Next, as we noted in Paper 1, we saw a possible increase in the rotation period during our 1999 observations which could only occur for the prograde case. However, the changing rotation period from apparition to apparition is in the same direction and therefore we cannot separate out the two effects. Finally, the morphology of the corkscrew we observed in 2010 September (discussed in the next section) favors the prograde solution, but the signal-to-noise of our images is too poor to be conclusive. Thus, it is likely that the prograde solution is the correct sense of rotation, but it is not yet definitive.

\section{DUST AND GAS MORPHOLOGY}
\label{sec:morphology}
Up until this point we have been concerned only with the nucleus brightness of Tempel 2. We turn now to an analysis of the dust and gas coma morphology. Previous authors have reported a persistent dust feature (cf. \citealt{sekanina79,sekanina87b}, and references therein). The radial nature of this feature implies a near-polar source which can be used to determine the pole orientation.  We are unaware of any previous reports of gas features. Thus, we discuss below the gas and dust morphology in our 1999 and 2010 data, and use them to constrain the current location of the pole.


\subsection{Morphology}
\label{sec:morph}
\begin{figure}
  \centering
  \epsscale{0.8}
  \plotone{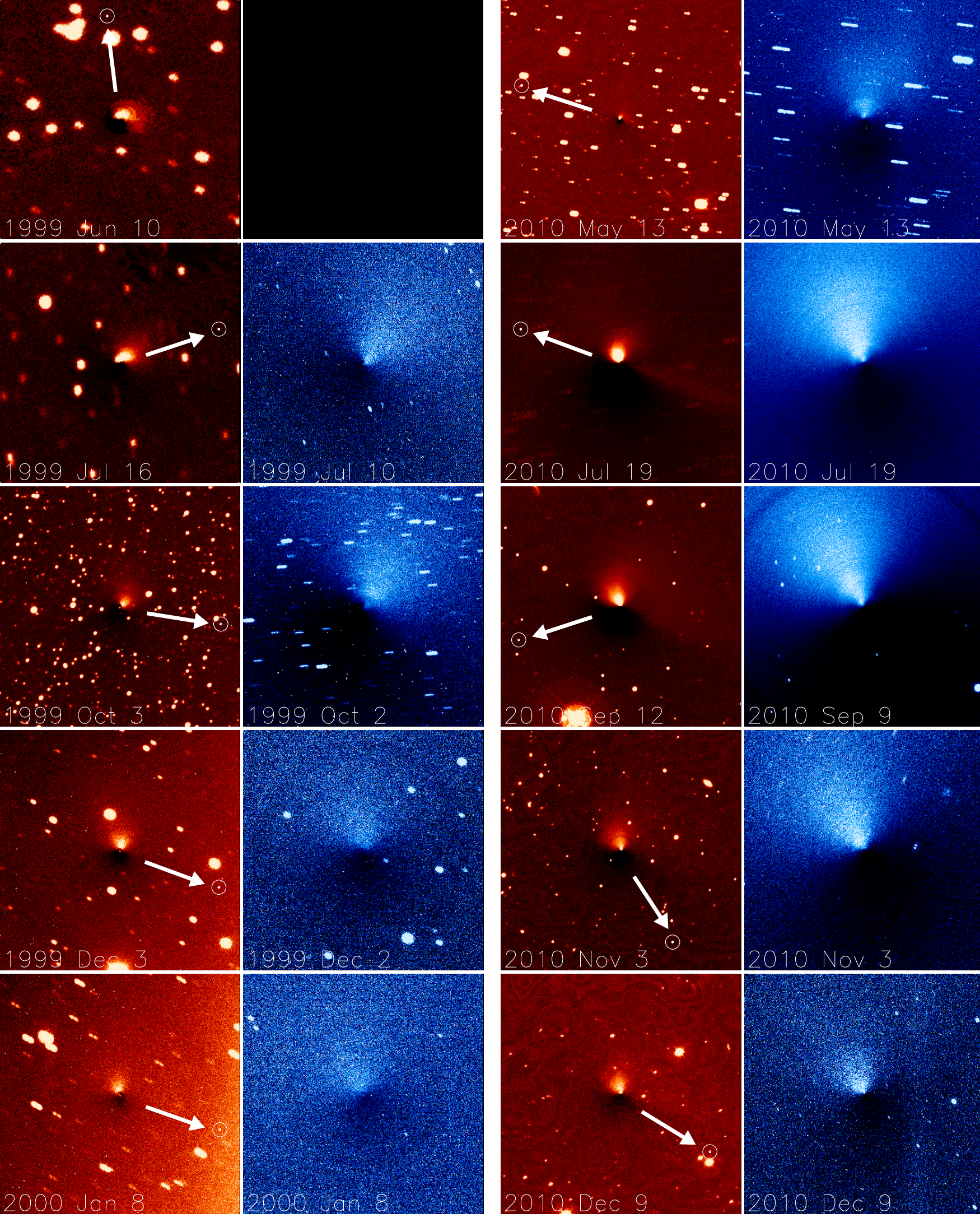}      
  \caption[CN and dust coma morphology]{Representative enhanced R and CN images at intervals during each apparition. Columns 1 and 2 are the 1999 R and CN images, respectively. Columns 3 and 4 are the 2010 R and CN images, respectively. The date of each image is given in the bottom left corner. Each image is centered on the nucleus with north up and east to the left, and has been enhanced by division of an azimuthal median profile. Trailed stars are evident in nearly all frames and dust along the comet's orbit is visible as a faint diagonal line running through the nucleus from the northeast to the southwest in the 2010 July 19 R image. No usable CN image was obtained during 1999 June, therefore this frame is intentionally left blank. The direction to the Sun is labeled for each pair in the R image and the phase angle for each night is listed in Table~\ref{t:imaging_circ}. The image widths are as follows: 76,000 km in 1999 June, 68,000 km in 1999 July, 330,000 km in 1999 October, 530,000 km in 1999 December, 680,000 km in 2000 January, 470,000 km in 2010 May, 300,000 km in 2010 July, 290,000 km in 2010 September, 430,000 km in 2010 November, and 630,000 km in 2010 December.}
  \label{fig:morphology}
\end{figure}

We enhanced the contrast of the morphological features by removing an azimuthal median (we used both division and subtraction at various times). These image enhancement techniques are relatively benign and are regularly used for morphological studies of comets (cf. \citealt{schleicher03a,schleicher04,lederer09,samarasinha11}). The technique involves centroiding on the comet, determining the median brightness as a function of distance from the nucleus, then dividing or subtracting the median value in each annulus from the original image. The result is a removal of the bulk radial profile, enhancing any azimuthal asymmetries while not altering the positions of such features. It is then much easier to analyze the locations and shapes of features in these enhanced images than in the original, unenhanced frames. We also investigated other enhancement techniques such as those discussed in \citet{schleicher03b} and \citet{farnham05}; the features discussed here were visible with a number of different enhancement techniques and in the raw images.
Figure~\ref{fig:morphology} shows representative enhanced R (columns 1 and 3) and CN (columns 2 and 4) images throughout both the 1999 and 2010 apparitions. 


The vast majority of our images of Tempel 2 during both apparitions were acquired with a broadband R filter since we were attempting to measure the rotation period via monitoring of the nucleus lightcurve. We acquired a handful of narrowband continuum (blue, green, and/or red) images on photometric nights in order to properly decontaminate the narrowband gas images (discussed below). All of these images looked similar to the broadband R images, suggesting that the broadband images are not strongly contaminated by gas (capturing the continuum over $\sim$250 nm dominates over any narrow gas features, even when the dust-to-gas is low). The narrowband continuum images had poorer signal-to-noise than the broadband R images so they are not discussed further. Tempel 2 has little dust and even when it was most active, the R images were dominated by a central condensation, although a faint dust tail and the fan-shaped dust jet described by previous authors was also evident. The dust jet was clearly distinct from the dust tail and was perpendicular to the anti-solar direction at times. The jet looked linear near the nucleus, but exhibited some curvature toward the tailward direction at larger distances. In the highest signal-to-noise images (e.g. 2010 July through September), the well defined linear portion of the jet was visible out to $\sim$50,000 km. The morphology looked constant throughout a rotation cycle, and no hints of any spiral shaped feature were seen within the jet. The position angle of the jet varied over the apparition as the viewing geometry changed. The jet's appearance was qualitatively similar throughout both the 1999 and 2010 apparitions. The radial appearance of the jet implies that it is located near the rotation pole and, as will be discussed in the next section, allowed us to analytically determine the orientation of the rotation pole.

As we have discussed elsewhere (e.g., \citealt{woodney02,knight12a}), CN is the most common gas species used to study coma morphology because it has a better contrast relative to dust than C$_2$, C$_3$, or NH and is much brighter than OH and NH. Since CN is minimally contaminated by underlying continuum and Tempel 2 is relatively dust poor, useful data can be obtained even under non-photometric conditions. The other gas species require removal of underlying continuum and, therefore, photometric conditions. Thus, we observed CN on both photometric and nonphotometric nights from 1999 June through 2000 January and 2010 May through December. We observed C$_3$ on photometric nights in 2010 August and September (it was not successfully observed in 1999) and C$_2$ on photometric nights in 1999 June through 2000 January  and 2010 September. We did not successfully observe OH or NH owing to Tempel 2's faintness.

As shown in Figure~\ref{fig:morphology}, CN was always enhanced in one hemisphere and had a nearly linear jet-like feature leaving the nucleus generally toward the northern hemisphere. The jet was visible much farther from the nucleus (in excess of 150,000 km in 2010 July through September) than was the dust feature. Similar to the dust, the position angle of the CN jet varied slowly over an apparition as the viewing geometry changed. There was minimal change in the CN coma morphology during a night and from night to night during a run, however, we see evidence of a faint corkscrew spiral within the jet in the highest signal-to-noise images in 2010. The signal-to-noise and number of hours per night available post-perihelion were both worse in 1999, preventing the discernment of comparable features. On nights with sufficient temporal coverage, we could discern outward motion of the turns of the corkscrew. From this motion we estimated the projected gas outflow velocity to be 0.6 $\pm$ 0.2 km s$^{-1}$ and used it as a constraint in our modeling (Section~\ref{sec:modeling}).

We typically acquired 1--3 images each of C$_3$ and C$_2$ on photometric nights. To the extent comparison with CN was possible given the generally poorer signal-to-noise, the C$_3$ and C$_2$ coma morphology agreed with the CN coma morphology. The C$_3$ and C$_2$ brightness was enhanced throughout the same hemisphere and with approximately the same central position angle as the CN jet. What differences were discernible between CN, C$_3$, and C$_2$ can be attributed to their differing lifetimes and parentages; C$_3$ did not extend as far as either CN or C$_2$ owing to the shorter lifetimes of it and its parents (cf. \citealt{ahearn95}), while C$_2$ extended roughly as far as CN but was more diffuse, consistent with its having multiple parents and grandparents (as compared to CN which is primarily a daughter species). We did not obtain enough images of C$_3$ or C$_2$ on a given night to assess rotational variability, but given their similarities to CN whenever each was observed, we infer that neither the C$_3$ nor C$_2$ varied substantially during a rotation cycle. Combined with the fact that they exhibit similar production curves (Section~\ref{sec:photometry}), we conclude that CN, C$_2$, and C$_3$ all likely derived from the same source region(s) on the nucleus. 

\subsection{Pole Solutions}
\label{sec:pole_soln}
The dominant coma feature throughout both the 1999 and 2010 apparitions is a nearly radial jet that shows no change with rotational phase in R and minimal changes in CN images. This implies that the jet's source region is located near the rotational pole of the nucleus and that the position angle (PA) of the middle of the observed jet corresponds to the projected direction of the pole on the plane of the sky because the corkscrew that would be produced by a source near the pole should have a center at the pole. Thus, by measuring the PA of the jet we can determine a great circle that defines the orientation of the comet's nucleus for each observation. As the observing geometry changes, different great circles are defined, and the intersection of these circles yields the orientation of the comet's pole.

\begin{figure}
  \figurenum{4}  
  \centering
  \vspace{-0.3in}
  \includegraphics[width=88mm]{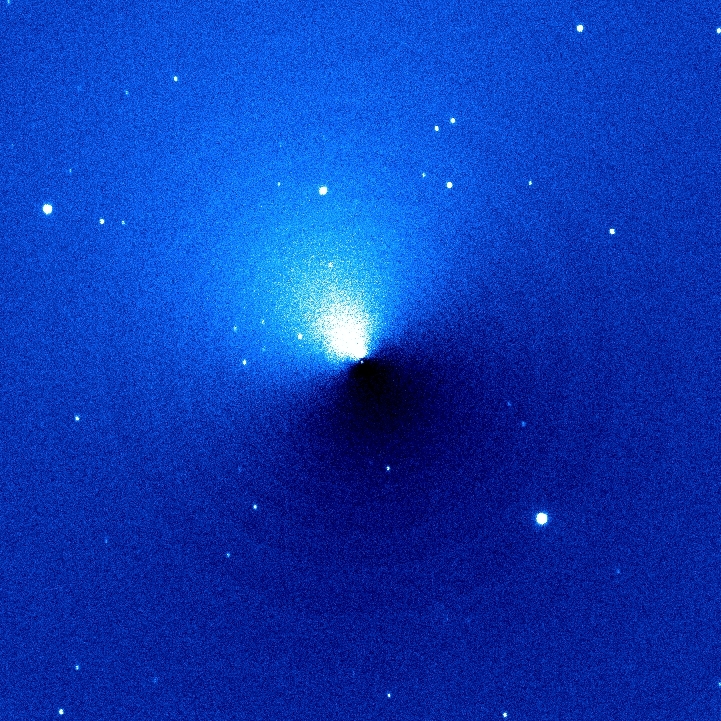}\\
  \vspace{-0.15in}
  \phantom{0}  
  \includegraphics[width=88mm]{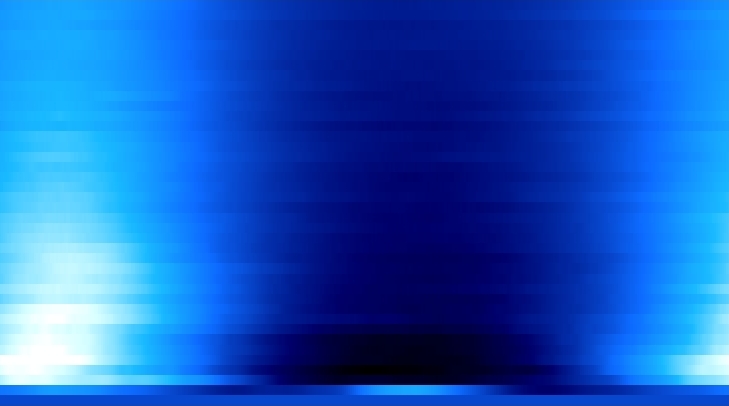}\\
  \vspace{-0.15in}
  \phantom{0}  
  \includegraphics[width=88mm]{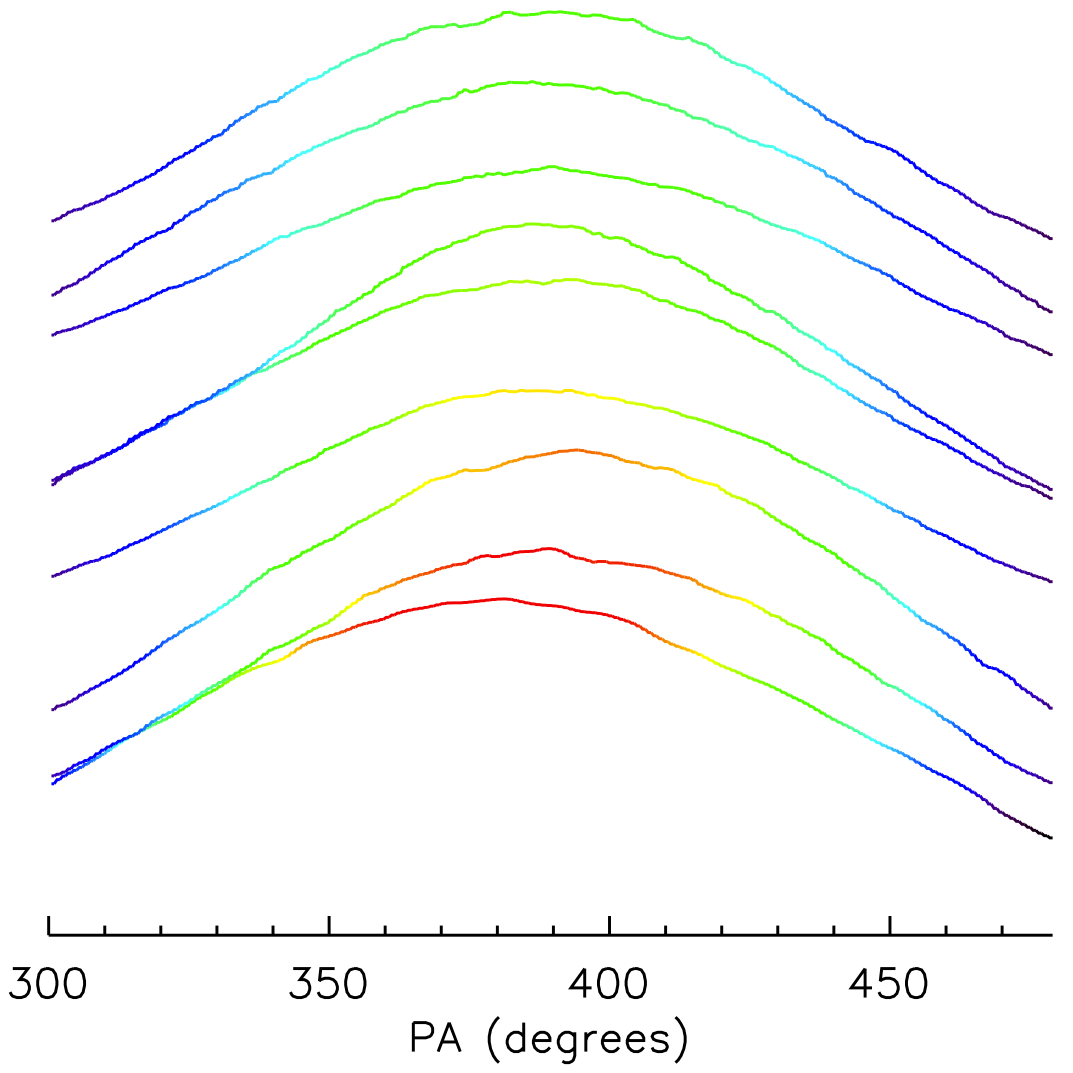}  
  \vspace{-0.15in}

  \label{fig:unwrap}
\end{figure}

In order to measure the PA of the jet, we started with an enhanced image centered on the nucleus then ``unwrapped'' it into $\rho$ (radial distance from the nucleus) and $\theta$ (azimuthal angle, with 0$^\circ$ to the north and measured counterclockwise through east) coordinates and smoothed it in $\theta$. This has the advantage of increasing the number of pixels in a given range of $\theta$ as $\rho$ increases, effectively improving the signal-to-noise as the jet gets fainter. 
Then we plotted each $\rho$ as a function of $\theta$ in a profile and measured the PA of the jet feature as a function of distance from the nucleus. 
A similar technique has been used by ourselves and others to measure the PA of radial features in comet comae, e.g., 19P/Borrelly \citep{schleicher03b} and 103P/Hartley 2 \citep{mueller12}. We illustrate this process in Figure~\ref{fig:unwrap}, where the top panel shows an image from 2010 September 9 enhanced by subtraction of an azimuthal median profile, the middle panel shows the same image ``unwrapped'' into $\rho$-$\theta$ coordinates and smoothed, and the bottom panel shows profiles of the subsection of the image centered on the jet. 


We measured the jet's PA at up to 10 equally spaced binned distances, $\rho$, in selected R and CN images each night. The step size varied from 3,000--20,000 km per bin depending on both the brightness of the feature and the geocentric distance.
In general, shorter maximum radial ``unwrap'' distances were required for the dust images due to the fact that the dust brightness falls off much faster than the gas. We typically did not measure the first two $\rho$ profiles as uncertainties in the centroiding (as discussed in Section~\ref{sec:ccd_reductions}) can be misleading at small distances where each pixel subtends a very large angle $\theta$.
We measured fewer rows early and late in the apparitions, when the comet was fainter and the jet was not detectable as far from the nucleus. We excluded rows where the measured jet position was clearly influenced by stars or other anomalies and, for the R images, we excluded rows where the effects of radiation pressure became evident. 
Separate nightly PAs and corresponding uncertainties were calculated for the CN and R images, and are given in  Table~\ref{t:jet_pas}.

\begin{figure}
  \figurenum{5} 
  \centering
  \vspace{-0.3in}
  \plotone{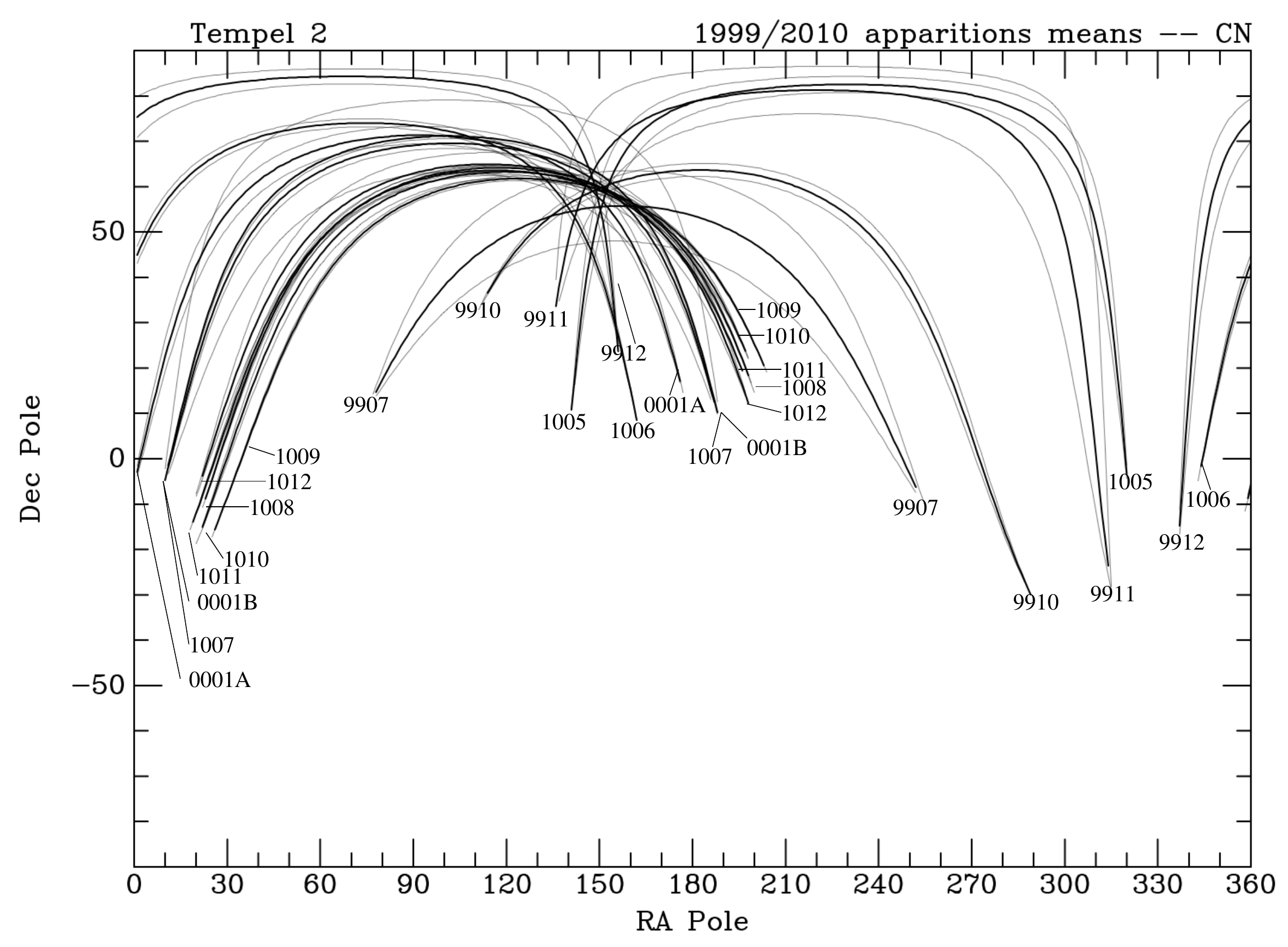}\\  
  \vspace{-0.2in}
  \phantom{0}  
  \plotone{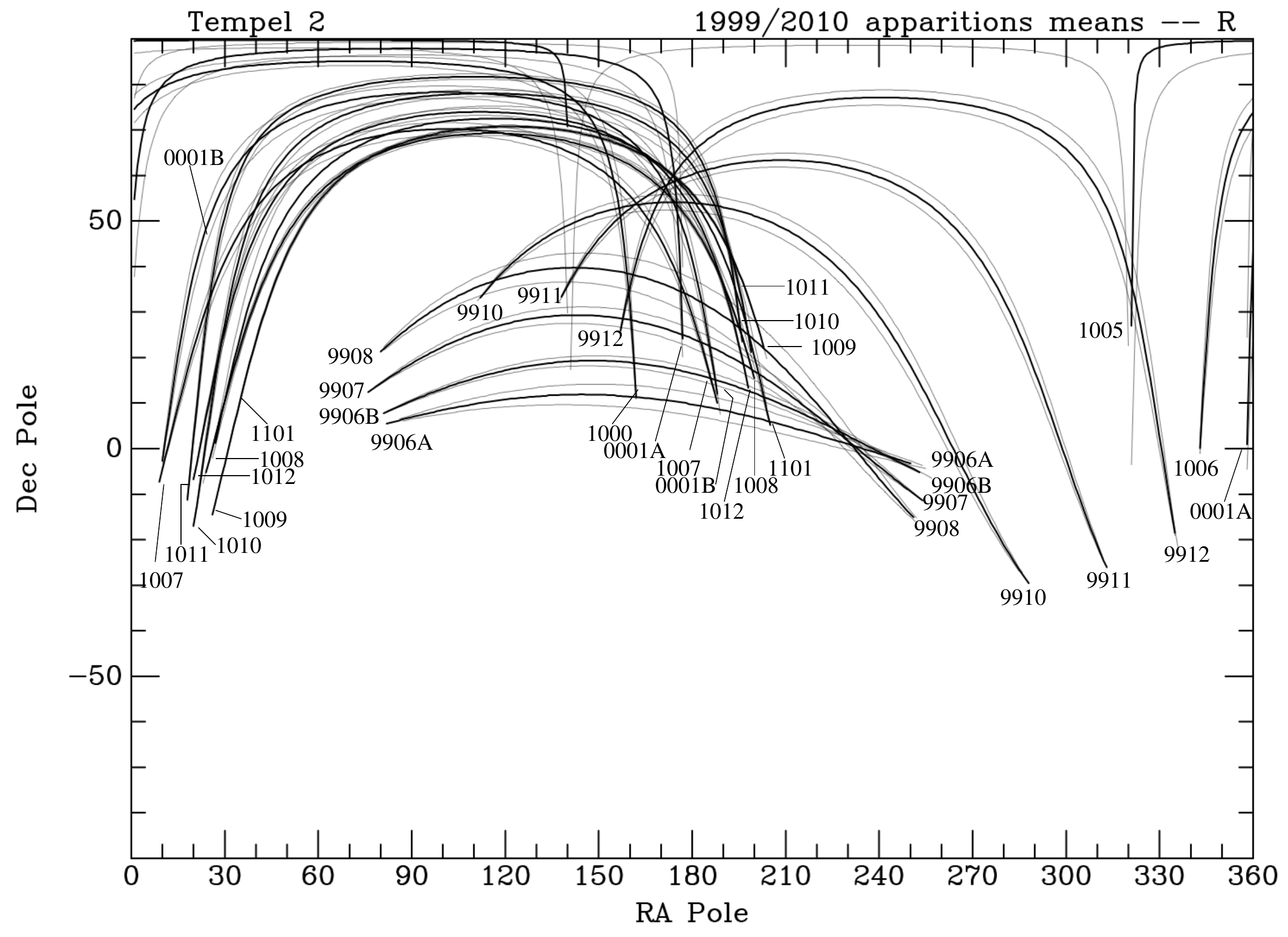}\\  
  \vspace{-0.0in}
  \caption[Pole Solutions]{Pole solutions from the 1999 and 2010 data for CN (top) and R (bottom). Each thick line is the great circle defining possible pole solutions for the mean PA of the feature over an observing run. The thinner lines on either side of the thick line are the uncertainty in the solution. Nightly PA measurements and their uncertainties are given in Table~\ref{t:jet_pas}. For runs in which a PA was measured on more than one night, the weighted mean PA was used. Each curve is labeled with the two-digit year and two-digit month with an additional A or B used in months with two separate runs (i.e., 0001A represents data from 2000 January 6--8).}
  \label{fig:pole_solutions}
\end{figure}

We plot the great circles corresponding to the PA measurements in Figure~\ref{fig:pole_solutions} for CN (top) and dust (bottom). The intersection of these great circles gives an analytical pole solution.
Our preferred CN pole solution is RA/Dec = 151$^\circ$/$+$59$^\circ$ (all RA/Dec coordinates are given in J2000.0) with a circular uncertainty of radius 3{\deg} in spherical coordinates. Nearly all of the individual great circles pass through this position and, within the uncertainties, we found the same pole solution when considering the 1999 (149$^\circ$/$+$61$^\circ$) and 2010 (153$^\circ$/$+$58$^\circ$) apparitions separately.  


The pole solution for dust is much less straight forward. Our preferred dust pole solution from both apparitions is RA/Dec = 173$^\circ$/$+$57$^\circ$ with a circular uncertainty of radius 4{\deg}, very similar to our preferred pole solution from the 1999 apparition (RA/Dec = 172$^\circ$/$+$56$^\circ$). However, there were a number of great circles which deviated significantly from this solution and require explanations. The earliest curves from the 1999 apparition (June through August) are too low and intersect near RA/Dec = 230$^\circ$/$+$2$^\circ$.
The curves deviate progressively less as the time from perihelion decreases and the behavior can be explained as the transition from a coma dominated by the large dust grains in the comet's trail (which are independent of the polar orientation) to one dominated by the increasing activity of the polar jet. As will be shown in the next section, Tempel 2 turns on rapidly 60--90 days before perihelion ($\Delta$$T$ = $-$60 to $-$90 days). Our earliest dust PA measurements (1999 June) occurred at $\Delta$$T$ = $-$89 days and matched the expected PA of very old material in the comet's orbit. Thus, what appeared to be the same dust jet as seen later in the apparition was actually the projection of the much older dust trail.
During 1999 July and August, our measured PAs gradually moved closer to the PA of the pole, 
suggesting that as the jet brightened, our measurements were influenced progressively more by the jet and less by the trail. We verified this explanation by examining published PA measurements of the dust jet from the 1988 apparition, when the viewing geometry was nearly identical to 1999. The earliest data (1988 May 17, $\Delta$$T$ = $-$123; \citealt{west88}) have a PA near the predicted position of the trail and more than 60{\deg} from the predicted PA of our pole. The PAs from data collected in June ($\Delta$$T$ = $-$87 days; \citealt{jewitt89}) and July ($\Delta$$T$ = $-$55 days; \citealt{boehnhardt90}) are intermediate between the expected PAs of the trail and of our pole, while the PAs from August through November \citep{boehnhardt90,campins90a} agree well with our predicted pole PA and not with the PA of the dust trail. Now turning to the 2010 apparition, as shown in Figure~\ref{fig:pole_solutions}, most of the curves are nearly parallel due to unfortunate viewing geometry, making it difficult to determine a unique dust pole solution for the apparition. However, the majority of curves cross the 1999 curves in the vicinity of the 1999 pole solution, confirming the solution. The exceptions were the set of curves which cluster near RA/Dec = 192$^\circ$/$+$42$^\circ$,  from 2010 October through December. During this time, the viewing geometry changed significantly and the dust tail swung rapidly from the northwest to the northeast, passing through the projected direction of the dust jet. Thus, our dust jet measurements during these months were likely influenced by contamination from the dust tail. 

The offset between pole solutions based on the dust and the CN morphology 
is both interesting and an annoyance, because there are several possible 
causes that are difficult to distinguish among. For instance, the dust 
feature might not be centered along the rotation axis because of 
radiation pressure even though this effect would be expected to be very 
small at our spatial scales. Evidence for this at larger projected 
distances from the nucleus was in fact a constraint for where we placed 
the limit on our measurements. 
As already noted, while a dust tail is 
sometimes evident in Tempel 2, we can usually clearly distinguish it from 
the dust jet. However, the tail does affect the background, introducing 
additional asymmetries in the coma that are not removed in the image 
enhancements. 
Dust release from the nucleus is also likely to exhibit a diurnal variation 
peaking at or shortly after local noon, assuming the source is not exactly 
at the pole. Note that radiation pressure would yield a shift in the 
location of the dust feature anti-sunward while any diurnal effect 
would be sunward. CN would also be expected to exhibit some 
diurnal variations in production, again in the sunward direction, but 
our data on other comets suggest the effect would be much smaller than 
for dust -- dust appears to need a threshold of gas density to be lifted 
which is not reached at night even though gas vaporization continues at 
a lower level. Nearly all of these effects will be modulated by the phase angle. When the phase angle is small (i.e., the Sun is nearly behind the observer) these effects will largely be along the line of sight and will thus appear compressed, while a large phase angle (i.e., the Sun is nearly perpendicular to the observer-comet line) will result in larger effects in the plane of the sky.
A more subtle process can also affect the observed location of the 
peak brightness in both dust and CN. As discovered by \citet{samarasinha00}, 
when jets broaden in width, the 
brightest location in a jet may no longer be the center of the jet 
due to projection effects. Since CN is a daughter product and has 
additional dispersion due to the excess velocity of its parent's 
dissociation, this effect would be expected to be larger for CN than 
for the dust. 

From checking behaviors throughout both apparitions,
we can identify no single phenomenon that consistently explains the CN--R 
offset. The observed morphology is likely due to differing combinations 
of phenomena for both CN and dust. Thus, 
we can not simply claim that the 
derived pole solution from CN is correct and the dust solution is 
offset for these reasons, nor can we claim the opposite. Without a 
clear preference, we have therefore chosen to split the difference 
and use the intermediate location (RA/Dec = 162{\deg} $\pm$ 11{\deg}/$+$58{\deg}
$\pm$ 1{\deg}) as our preferred solution. This necessitates an 
elongated region of uncertainty in order to encompass both the CN and 
dust pole solutions. It is possible that the orientation of the
rotation pole is changing with time (and is thus responsible for the
change of rotation period). We do not see any evidence for a change 
in the pole, but any such change is likely to be within our
quoted uncertainty.


\citet{sekanina87b} collected historical observations of a ``fan-shaped coma'' exhibited by Tempel 2, finding a pole oriented at RA/Dec = 147.9$^\circ$/$+$54.8$^\circ$  from 1925--1967 (where we have converted his B1950.0 solution to J2000.0). Later, \citet{sekanina91} derived a solution using only data from 1988, finding RA/Dec = 139.9$^\circ$/$+$48.8$^\circ$ (again, we have have converted his B1950.0 solution to J2000.0). These solutions are near our own combined solution, differing by 8.8$^\circ$ and 16.2$^\circ$, respectively. Possible reasons for differences in our pole solutions include inconsistencies in what was measured (such as possible confusion with a dust tail and/or a second jet), due to the numerous methods undoubtedly utilized by the various authors whose measurements \citet{sekanina87b} used (Sekanina himself excluded all measurements from 1899, 1920, and 1972), the short baseline of data used in 1988 (Sekanina stated that the difference in his solutions was not ``compelling enough to prefer this position of the pole to the `standard' one [from 1925--1967]''), and simplifying assumptions made by Sekanina such as that the comet's rotation pole must be centered at the middle of the jet's axis or that the gas coma would be radially symmetric (he noted that the blue sensitivity of the emulsion on plates from 1920 and 1925 would lead to contamination by CN). It is unclear whether differences between Sekanina's 1925--1967 pole and our own is due to a precession of the pole or due to one or more of the differences discussed above. As will be seen in Section~\ref{sec:water_prod}, our pole solution naturally explains the pre-/post-perihelion asymmetry better than either of these earlier solutions.



\subsection{Numerical Modeling}
\label{sec:modeling}
Having determined a pole solution analytically, we use our Monte Carlo jet model, described in detail in \citet{schleicher03a} and \citet{schleicher03b}, to reproduce the coma morphology. The model allows us to vary (among other things) the time, viewing geometry, location and extent of the active region, outflow velocity, and the amount of dispersion from the local normal direction, and it accounts for the dynamics of both gas and dust after they leave the nucleus. Since we have already established that the jet must be located near the pole, the primary guide for our modeling was the faint CN corkscrew seen in 2010 September. 

We investigated a small range of latitudes near the pole (varying the longitude only shifts the timing of the corkscrew). We found that a source at latitude $+$80{\deg} best reproduced the angular extent of the corkscrew; sources at $+$75{\deg} and $+$85{\deg} fared noticeably worse. Testing revealed that the extent of the source made little difference. However, as will be discussed in Section~\ref{sec:water_prod}, we found that a source of radius $\sim$10{\deg} best reproduced the measured water production rate. We then varied the outflow velocity until the spacing and motion of the turns in the corkscrew approximately matched our images. This yielded an outflow velocity of 0.7 $\pm$ 0.2 km s$^{-1}$, about 20\% higher than the projected velocity we estimated from the images. The model naturally revealed why we did not observe a corresponding corkscrew in dust images. Since the velocity of dust is much lower than it is for gas (typically 1/3 to 1/2 of the gas velocity), the spacing in the dust corkscrew is much closer together than in the CN corkscrew. Furthermore, there is considerable velocity dispersion in the dust grains, resulting in a smearing out of the corkscrew shape. The combined effect is that the gaps between turns in the corkscrew are filled in and the dust jet looks like a cone instead of a corkscrew. 



\section{NARROWBAND PHOTOMETRY}
\label{sec:photometry}

\subsection{The Photometry Data Set}
We now turn to the narrowband photoelectric photometry data, from 
which we determine gas and dust fluxes, and derive gas production 
rates and dust {\afrho} values. 
Emission band and continuum fluxes are listed in Table~\ref{t:phot_flux}. 
Applying 
the nightly fluorescence efficiencies that are functions of heliocentric 
distance and velocity (Table~\ref{t:phot_circ}), we convert the gas fluxes 
to column 
abundances that are also listed in Table~\ref{t:phot_flux}. A standard 
Haser model 
is then applied, using the scalelengths given in \citet{ahearn95}, 
to extrapolate the column abundances to total coma abundances, and 
gas production rates ({\it Q}) are then computed by dividing the total 
abundances by the assumed daughter lifetimes; these are tabulated in 
Table~\ref{t:phot_rates}. As the only definitive parent species and also the most 
abundant volatile species, water production rates are derived based 
on its daughter, OH, and our empirical conversion from {\it Q}(OH; Haser model) 
to {\it Q}(H$_2$O; vectorial) (\citealt{cochran93}; also see 
\citealt{schleicher98}), and these are given in the last column of 
Table~\ref{t:phot_rates}.
Our proxy for dust 
production, \afrho, is also listed in Table~\ref{t:phot_rates} for 
each continuum 
filter. This quantity would be the same at each wavelength if the dust 
grains are gray in color and the same for differing aperture sizes if 
the dust follows a canonical 1/$\rho$ radial distribution.

Finally, the 1-$\sigma$ uncertainties based on the photon statistics 
associated with each data point are given following the $Q$s and 
\afrho's in Table~\ref{t:phot_rates}. Since such uncertainties are balanced in 
linear space (i.e. the same percentage $+$ and $-$), the logarithms will 
be unbalanced; for clarity, we only list the ``$+$'' log value. In some 
cases, 1-$\sigma$ upper limits for the flux and subsequent variables 
are given when the values were negative following the subtraction of 
sky and continuum. In the lone case where even the 1-$\sigma$ upper limit 
is negative, the result is listed as undefined.



\subsection{Temporal Behavior}
\begin{figure}
  \figurenum{6} 
  \centering
  \plotone{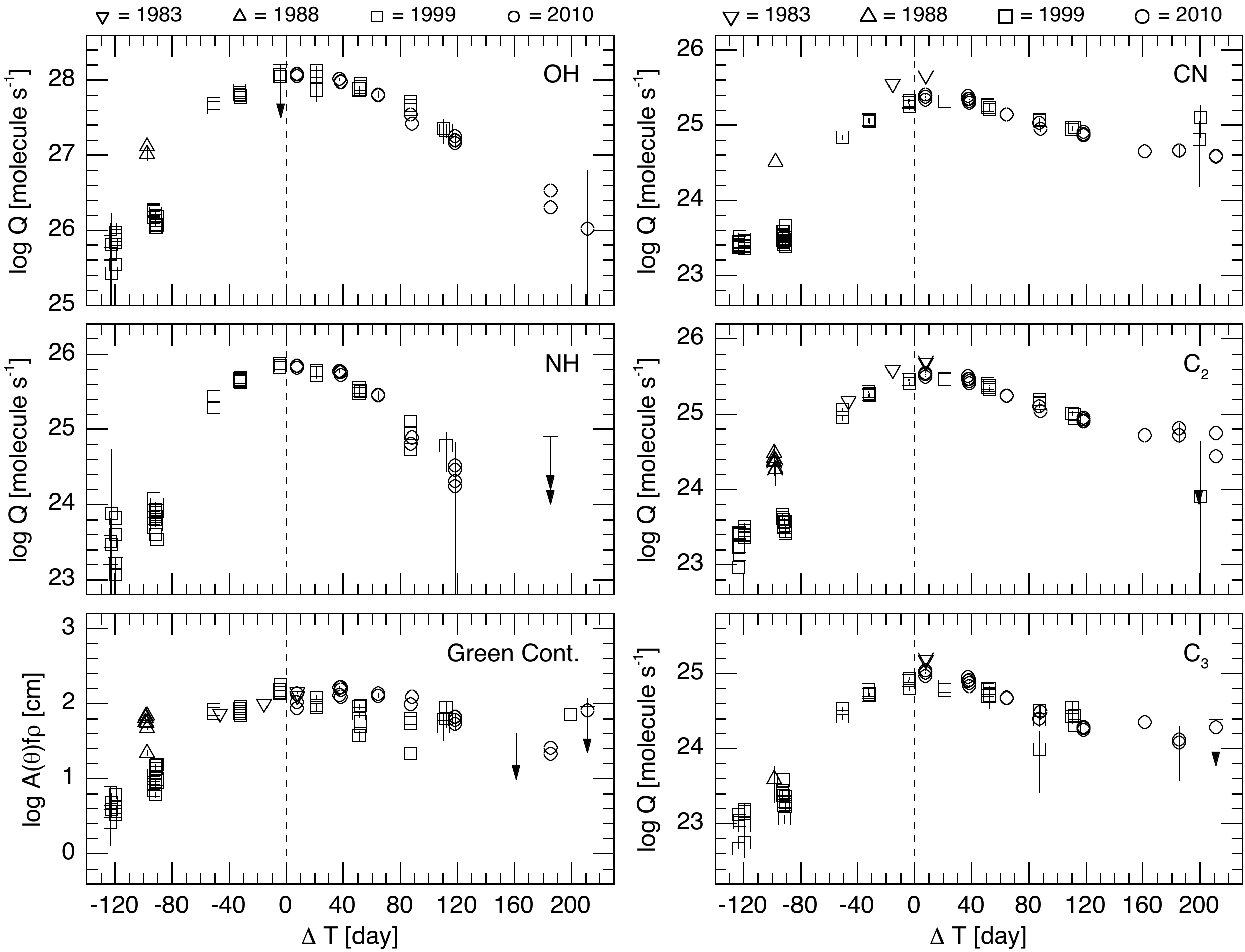}\\
  \caption[Gas and dust production rates]{Log of the production rates for each observed molecular species
and \afrho\ for the green continuum plotted as a function of time from 
perihelion. Data points from the 1983 apparition are shown as inverted 
triangles while those from 1988 are shown as triangles, those from 1999/2000
are given as squares, and the recent 2010/2011 data are shown as circles. 
Note the large asymmetries around perihelion for all species. The larger 
values in 1983 as compared to 1999 and 2010 are due, at least in part, to 
a smaller perihelion distance. While this also affects the 1988 data points, 
when the perihelion distance matched that of 1983, the large offset 
between 1988 and 1999 appears mostly due to a somewhat earlier ``turn-on'' 
of major activity in 1988.
Also note that the peak in production rates is unusually close to perihelion 
(between 0 and $+$20 days) for a comet exhibiting such a large asymmetry 
in activity.}
  \label{fig:phot1}
\end{figure}

The derived production rates are plotted as logarithms in Figure~\ref{fig:phot1}
as a function of time from perihelion. Immediately evident is the 
strongly asymmetric behavior before and after perihelion, with 
a large discontinuity about 3 months before perihelion, and a 
relatively smooth curve after. It is also evident that the 1999 and 
2010 apparitions are in excellent agreement even with a change in 
perihelion distance from 1.48 to 1.42 AU; the only discrepancies are 
small variations in {\afrho} later than about $\Delta$$T$ = 
$+$60 days (60 days after perihelion) and we have 
confirmed that these are caused by the difference in phase angles 
along with aperture effects. As in most comets, the radial profiles 
of the dust fall-off more steeply than the canonical 1/$\rho$ distribution, 
resulting in smaller {\afrho} values at progressively larger 
apertures. In comparison, the carbon-bearing gas species all exhibit 
small aperture trends opposite that of dust, while OH and NH show 
no consistent trends with aperture size. 


In contrast to the agreement between the 1999 and 2010 apparitions, 
the few 1983 observations near perihelion are systematically higher 
(by 60--90\%) for the measured gas species. While the perihelion 
distance is smaller, with $q$ = 1.38 AU, this difference in $q$ is less
than that between 1999 and 2010.
Based on our data alone, therefore, it is unclear whether or not a 
long-term secular change in Tempel 2's activity level is taking place; 
we will return to this issue in Section~\ref{sec:water_prod}. 

One definitive difference that we detect is the time of the ``turn-on'' 
of activity and the strong brightening of the coma. The two nights 
of observations in 1988 ($\Delta$$T$ $\sim$ $-$98 days) have production rates 
consistent with the post-turn on activity trend in 1999, while the 1999 
turn-on clearly did not occur until after $\Delta$$T$ $\sim$ $-$90 days. 
Absolute magnitudes from 
throughout the 1900s suggest that the time of significant activity turn-on 
apparently varies by a month or more with no discernable trend or pattern 
(cf. \citealt{sekanina79,boehnhardt90,churyumov92}).

We note that even when one excludes the pre-turn-on data, the 
$r_\mathrm{H}$-dependencies (Table~\ref{t:abundance_ratios}) are 
extremely steep before perihelion, with power-law slopes for the 
gas species and of the phase-adjusted dust of between $-$16 and $-$20. 
While this is over a very small range of heliocentric distance -- 1.57 to 
1.48 AU -- note that had we included our earliest data (1.9 AU) the 
slopes would have been even steeper. 
This implies that the level of activity continues to rapidly ramp-up 
after the near step-function ``turn-on'' already described. 
When the comet is outbound, the slopes are much more ``normal'' but 
also more diverse. For instance, the carbon-bearing species have 
slopes between $-$3 and $-$5, but there is curvature and the slopes are 
somewhat steeper just after perihelion and shallower beyond 2 AU. 
OH is significantly steeper, at $-$9, and slightly steepens with distance -- 
opposite the behavior of the carbon-bearing species. At this time we 
have no explanation for the difference between the carbon-bearing species
and OH given their similarity pre-perihelion. NH is even steeper, 
at $-$13, and no data were successfully obtained beyond 2 AU. 
Finally, as usual for comets receding from the Sun, dust is least steep, 
in this case with a slope of $-$3 (after adjusting for phase angle). 
As we indicated in our work on Comets 19P/Borrelly \citep{schleicher03b} 
and 67P/Churyumov-Gerasimenko \citep{schleicher06b}, we consider 
the shallower $r_\mathrm{H}$-dependent slopes for dust as compared to gas species 
outbound to most likely be caused by the presence of heavier, 
very slow moving grains lifted off the surface near peak outgassing, 
i.e., near perihelion. Such grains would remain in our photometer 
entrance apertures much longer than smaller grains, and Tempel 2 appears 
to provide yet another example. 


Given the much steeper power-law $r_\mathrm{H}$-dependencies before perihelion as 
compared to after, an unexpected finding is that peak production 
takes place only 1--3 weeks following perihelion, since 
for several other comets strong asymmetries are associated with 
larger offsets in the time of peak production.
However, our result here is consistent with normalized brightness estimates
from earlier apparitions (cf. \citealt{sekanina79,boehnhardt90,churyumov92})
and the $IUE$ measurements in 1988 \citep{roettger90}.

\subsection{Compositional Results}
The abundance ratios, listed in Table~\ref{t:abundance_ratios}, based on the ratios of 
the various species' production rates, place Tempel 2 in the 
midst of what would be called normal comets. 
Looking specifically at the C$_2$-to-CN ratio, 
Tempel 2 is classified as ``typical'', the same as was determined 
from the few data points available at the time of our original database 
paper \citep{ahearn95} and consistent with our ongoing database analysis
\citep{schleicher10}. 
Tempel 2 was also observed spectroscopically in the 1980s by several 
groups -- \citet{newburn89}, \citet{cochran92}, \citet{fink96}, and \citet{fink09} -- without any claims that the 
comet's composition was out of the ordinary, and \citet{fink09} lists 
it in his own ``typical'' class. Because all groups use differing 
scalelengths in their respective reductions to production rates, 
and nothing unusual was reported, we have not attempted to make 
more detailed intercomparisons among the datasets of the abundance ratios. 

Several issues have an effect on the derived dust-to-gas ratio 
that we normally compute by simply ratioing the {\afrho} values to Q(OH). 
First, as already discussed, the $r_\mathrm{H}$-dependencies of OH and dust 
after perihelion are quite different, leading to a significant trend 
with distance. Second, the aperture trends seen in dust also 
change the resulting ratio, although by a smaller amount than 
heliocentric distance. Third, phase angle effects for the dust 
are non-negligible. Finally, dust measurements obtained before the 
coma substantially turns-on are significantly contaminated by 
light reflected by the nucleus. Ultimately, we chose to avoid the 
contamination issue by excluding pre-turn-on data from the averages, 
while the trends with distance and aperture are reflected in the 
computed standard deviations.  The resulting value of log 
(\afrho/{\it Q}(OH)) was $-$25.69 $\pm$ 0.06 using the green 
continuum. The corresponding average value when the individual dust 
measurements were adjusted (by the values in Table~\ref{t:phot_circ}) 
for phase angle to 0\deg\ was $-$25.43 $\pm$ 0.05. 
The latter value can be compared to our current database results 
\citep{schleicher10}, and Tempel 2's dust-to-gas ratio is at the 
average of all comets and of comets having similar perihelion distances. 

As evident from the mean values of the dust-to-gas ratio for each of the 
three continuum filters in Table~\ref{t:abundance_ratios}, there is almost no variation with 
wavelength. This is confirmed by the very small amount of reddening, 
only 7\% per 1000 \AA\ between the green and UV continuum points. 
This is somewhat smaller than the mean color of 19\% per 1000 \AA\ between 
5660 and 4400 \AA\ we compute using the better quality data by 
\citet{storrs92} and again excluding pre-turn-on data points.
Finally, we do not detect any trend in our own dust colors with aperture size.

\subsection{Water Production and Model Comparison}
\label{sec:water_prod}
\begin{figure}
  \figurenum{7} 
  \centering
  \plotone{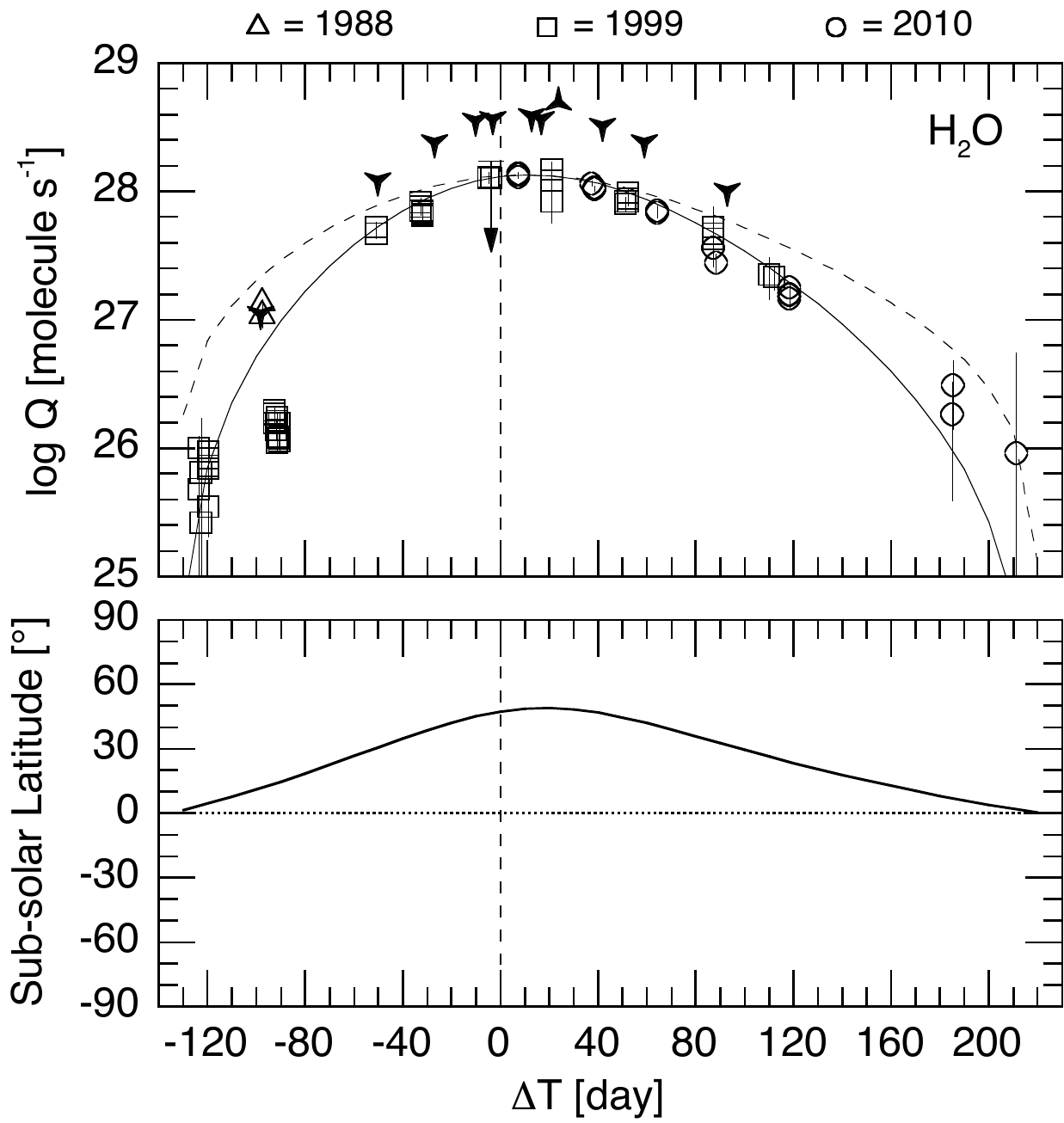}\\ 
  \caption[Log of water production rate relative to perihelion]{Log of the production rates for water plotted as a function of 
time from perihelion (top panel). Our values are shown with the same open 
symbols as in Figure~\ref{fig:phot1}, while results based on OH measurements with the $IUE$ 
satellite \citep{roettger90,feldman92b} are overlaid as ``filled Ys'' while a single 
datum based on a forbidden oxygen measurement \citep{fink09} is overlaid as 
an ``upside down filled Y''; these other data are each from the 1988 apparition. 
Note that our two nights of data in early 1988 are essentially identical 
to the first $IUE$ point. The clear offset in our 1999 and 2010 data 
in the months surrounding perihelion as compared to the 1988 data is 
similar to what we observed in 1983 in the minor species (see Figure~\ref{fig:phot1}), 
but the change in solar flux due to the change in perihelion distance is
too small to explain the offset.
The solid curve is the result of our coma jet model used to match 
the near-polar jet morphology observed in 1999 and 2010 (see Section~\ref{sec:modeling}), 
and represents our model solution for water vaporization originating 
primarily from the near polar source region and responding as the 
cosine-squared of the Sun angle. A canonical cosine of the Sun angle 
curve is shown as the dashed line and is clearly above the data beyond 
${\Delta}T$ = $\pm$80 days.
The sub-solar latitude from our model solution is plotted as a 
function of time (bottom panel).
}

  \label{fig:phot2}
  \label{lastfig}
\end{figure}

Using the $Q$(H$_2$O) values given in the last column of Table~\ref{t:phot_rates}, 
we plot in Figure~\ref{fig:phot2} water production as a function of time. 
These values can be compared to measurements of the same ultraviolet 
OH band using the $IUE$ spacecraft during the 1988 apparition 
\citep{roettger90,feldman92b}. A large and nearly 
constant offset of about a factor of 2.5, is present between 
the 1988 values and our 1999 and 2010 results. This offset is somewhat 
larger than the factor of 1.9 value for $Q$(CN) in 1983 as compared 
to the most recent apparitions;
note that 1983 and 1988 had 
the same perihelion distance, so the same cause might explain the 
observed offsets for both the water and the minor species. 
Also note that the only water measurements we obtained in 1988 
($\Delta$$T$ = $-$98 day) were concurrent with the first $IUE$ point and are 
essentially identical, implying the observed change between apparitions 
previously mentioned is not an artifact of modeling parameters. 
The only other published water measurement of which we are aware 
is a single value by \citet{fink09} in late 1988 based on 
an OI[1D] measurement; this result is about 40\% higher than the 
corresponding $IUE$ data.



After allowing for an {\rh}$^{-2.5}$ contribution due to the change in 
perihelion distance to the production rate changes between apparitions 
(20\% drop from 1988 to 1999; 12\% increase 
from 1999 to 2010), it is obvious that for the 1999 and 2010 
curves to be essentially identical implies about a 12\% secular decrease. 
This is far less than the 60\% decrease observed between 1988 and 1999, 
of which only one-third can be attributed to the change in perihelion 
distance, implying a 40\% secular drop. While our two OH data points 
in 1988 match the $IUE$ point, some of this secular drop might be an 
artifact.  We also note that there is no evidence for a very large 
decrease in activity over the last century, as implied if the decrease 
from 1988 to 1999 was ongoing. We therefore suspect that the variations 
from orbit to orbit are irregular, and are perhaps associated with 
the clumpiness of ice within the source region.

Our nucleus model can be used to compute the predicted water production 
rate in a manner similar to that used for Comet 19P/Borrelly 
\citep{schleicher03b}. The basic procedure is relatively simple, since the 
location of the rotational pole directly gives the sub-solar latitude 
as a function of time, and thereby the angle of the Sun from the zenith 
for a polar source. We initially assume that the amount of outgassing 
is proportional to the intensity of the sunlight, i.e., to the cosine 
of this Sun angle. While our preferred jet model has the source region 
centered at a latitude of about $+$80\deg\ with a 10\deg\ radius, rather than located directly 
at the pole, the rotationally averaged cosine is conveniently nearly 
identical to the polar solution, greatly simplifying the computations. 
The other component in the model is just the water vaporization model 
we generally use to compute the active area required to produce the 
observed water production at a given heliocentric distance. 
Here we continue to use a vaporization curve by A'Hearn\footnote{http://www.astro.umd.edu/$\sim$ma/evap/} 
based on the work of \citet{cowan79}, specifically the case 
for a pole-on, rapidly rotating nucleus. This model case assumes 
a uniformly active surface (modulated by the amount of sunlight) 
rather than a small, isolated source at this subsolar pole. The intensity 
is thus four times higher due to solar emission perpendicular to the surface 
as compared to averaged over a sphere, and so we adjust the nominal 
peak computed active area of 8.5 km$^2$ downward by a factor of four
to 2.1 km$^2$. However, since the subsolar latitude is actually
substantially lower than the pole (see discussion below), the intensity
is lower and therefore the size of the active area  must be increased.
Thus, we adjust the active area upward to about 3.7 km$^2$.
Note that, based on the effective
radius of 5.98 km \citep{lamy09}, the total surface area of Tempel 2 
is 450 km$^2$. This yields an active fractional area of about 0.8\%, 
corresponding to the effective radius of the source region of 10\deg\ 
that we employed in the jet model in Section~\ref{sec:modeling}. 

The derived sub-solar latitudes based on our pole solution are given in 
the bottom panel of Figure~\ref{fig:phot2}. The peak value of 49\deg\ is reached at 
$\Delta$$T$ = $+$19 days for the 2010 apparition; the time of the peak changes 
by less than 1 day with the variations in $q$ for the other apparitions. 
The resulting model water production, based on the cosine of the Sun 
angle combined with the water vaporization model with distance, 
directly yields a peak production matching that observed in 1999 and 2010. 
However, the model curve is too shallow as compared with the data 
at smaller sub-solar latitudes (i.e., larger Sun angles) as shown with 
the dashed curve in the top panel of Figure~\ref{fig:phot2}. 
One can ask if one or more additional source regions 
could steepen the curve, but a polar source already yields the 
steepest result, and any additional sources would instead make the 
total solution more shallow. We, instead, can obtain a much better fit 
by using a stronger function with solar angle, specifically a 
cosine-squared illumination function rather than the basic cosine 
function, shown in the top panel of Figure~\ref{fig:phot2} by the solid curve. 
Such a solution might imply that sunlight needs to penetrate into 
the surface of the nucleus and is much more efficient when the Sun 
is high in the sky as viewed from the source region. Alternatively,
this could be due to local topography that creates significant shadowing 
until the Sun is high in the sky. The fact that 
there is also strong evidence for a progressive secular decrease could
mean that the sunlight must penetrate deeper each apparition, but 
the $IUE$ data from 1988 do not suggest that the solar angle function 
was less steep in the past, so we are unwilling to speculate further 
regarding the details of the physical processes taking place at and 
just below the surface of Tempel 2 during vaporization. 

No simple function can explain the abrupt turn-on of activity observed 
by ourselves and others between $-$100 and $-$70 days, but in our model this 
corresponds to a sub-solar latitude of only 11--22{\deg}, suggesting that 
local topography and resulting shadowing likely dominates the process. 
Supporting this scenario is that the reverse circumstances take place 
between $+$125 and $+$165 days, just when the brightness has exhibited 
a significant drop at some apparitions (cf. \citealt{sekanina79}). 
Note that the model solution for the outgassing is quite sensitive 
to the specific pole solution.  In fact, our solution based on 
the directions of the CN jet alone yield a peak sub-solar latitude 
only 8 days after perihelion, resulting in too little 
pre-/post-perihelion asymmetry. Both pole solutions by Sekanina 
(\citeyear{sekanina87b}; \citeyear{sekanina91}) 
have a similar problem, with the earlier solution peaking at 
$+$6 days and the latter peaking at perihelion and resulting in no 
asymmetry at all. Could the asymmetry instead be due to large thermal lags 
rather than the axial tilt? We consider this unlikely, first because 
the timing of abrupt turn-on and turn-off of activity just discussed, 
and also based on the investigation by \citet{ahearn95} showing 
that equal numbers of comets are brighter before perihelion as after, 
which would not be true if thermal lags dominate.

\section{SUMMARY}
We have reported on new nucleus lightcurve data from the 2010 apparition, jet morphology from the 1999 and 2010 apparitions, and traditional photoelectric photometry from the 1983, 1988, 1999, and 2010 apparitions. All imaging and nearly all photometry were acquired at Lowell Observatory. The observing circumstances and observational goals were different during the two primary imaging apparitions. Opposition occurred prior to perihelion in 1999; coupled with the desire to measure a nucleus rotation period (reported on in Paper 1), our 1999 observations were more extensive prior to significant ``turn-on'' of activity and were heavily weighted toward broadband-R. Opposition occurred after perihelion in 2010, resulting in nearly all imaging data having been acquired after ``turn-on'' of activity. While we sought to measure the rotation period in 2010 for comparison with previous apparitions, we also looked to characterize the coma morphology in order to obtain a robust pole solution. Thus, fewer broadband-R and more narrowband gas images were acquired in 2010 as compared to 1999. 


Despite nearly all of our observations in 2010 having occurred after significant ``turn-on'' of activity, we successfully extracted a nucleus lightcurve and measured a nucleus rotation period of 8.950 $\pm$ 0.002 hr. This was longer than we previously measured for 1999 (8.941 $\pm$ 0.002 hr) and 1988 (8.932 $\pm$ 0.001 hr), confirming our prediction in Paper 1 based on the spin-down from 1988 to 1999 and the assumption that the change in period was caused by on-going torques. This makes Tempel 2 just the second comet (after 9P/Tempel 1; \citealt{belton11}) to have exhibited a change of rotation period on multiple orbits, and rules out scenarios for a one-time change in period such as impact or a fragmentation event. We used our pole solution to determine the sidereal period (8.947 hr for the prograde case, 8.953 hr for the retrograde case) and confirmed that the sidereal period had changed since 1999 (8.938/8.944 hr for prograde/retrograde) and 1988 (8.930/8.934 hr prograde/retrograde). Evidence suggests that the sense of rotation is prograde, but we cannot yet rule out the retrograde solution, and thus the precise sidereal period is still uncertain. Regardless of the sense of rotation, the change in period per perihelion passage from 1999 to 2010 (three intervening perihelia) was smaller than the change in period per perihelion passage from 1988 to 1999 (two intervening perihelia). It is unclear if this was due to the decrease in production since 1988, due to the 2010 observations being made before the end of outgassing (and thus before the full spin-down could occur), or simply a result of the uncertainty in the period measurements each orbit. The observed spin-down is consistent with that expected given Tempel 2's measured activity and inferred size.



We also investigated the dust and gas coma morphology in both 1999 and 2010. A nearly radial jet in both R and CN was observed that changed minimally (CN) or not at all (R) during a rotational cycle, implying that the jet's source region is located near the rotation pole of the nucleus. Hints of a corkscrew were seen in the CN jet in 2010 September. Despite low signal-to-noise in these images, we measured a gas projected outflow velocity of 0.6 $\pm$ 0.2 km s$^{-1}$ (and derived an actual outflow velocity of 0.7 $\pm$ 0.2 km s$^{-1}$). The coma morphology of C$_2$ and C$_3$ was consistent with the CN coma morphology after accounting for their varying parentages and lifetimes and, coupled with their similar production rate curves, implies they are all released from the same source region(s).

We measured the position angle (PA) of the center of the jet in both R and CN images throughout each apparition. Assuming that the center of the jet corresponds to the rotational pole of the nucleus, the intersection of the great circles representing each of these PAs defines the orientation of the rotational pole. Unexpectedly, the pole solutions differed for CN (RA/Dec = 151{\deg}/$+$59{\deg} with a circular uncertainty of 3{\deg}) and R (RA/Dec = 173{\deg}/$+$57{\deg} with a circular uncertainty of 4{\deg}). A number of factors were considered which could cause the center of the jet to appear offset from the pole, with our preferred solution (RA/Dec = 162{\deg} $\pm$ 11{\deg}/$+$58{\deg} $\pm$ 1{\deg}) ultimately chosen to be midway between the R and CN solutions. This pole is near those found previously by \citet{sekanina87b,sekanina91}, but does a better job of matching the pre-/post-perihelion asymmetry of the water production rate. We conducted Monte Carlo numerical modeling using this pole to replicate the CN corkscrew, finding that the morphology was best replicated for a source $\sim$10\deg\ in radius located near a comet latitude of $+$80{\deg}.


Photoelectric photometry was conducted to extensively measure the production rates of multiple species during the 1999 and 2010 apparitions, as well as much more limited observations during the 1983 and 1988 apparitions. The 1999 and 2010 data are in excellent agreement in spite of small changes in perihelion distance (1.48 AU in 1999 versus 1.42 AU in 2010). In contrast, the 1983 and 1988 data are significantly higher, CN by a factor of 1.9 and water by a factor of 2.5 (using $IUE$ data which were consistent with our limited OH data), even though the perihelion distance in 1983 and 1988 was only somewhat smaller (1.38 AU) than in 2010. The difference in production rate is much larger than would be expected for a simple change in insolation. As there is no evidence that Tempel 2 was fading steadily throughout the 20th century, we suggest that the decrease is not secular but is instead due to variations in the topography and clumpiness of ice in the primary source region.


The exact timing of ``turn-on'' varied, but occurred around three months prior to perihelion. The {\rh}-dependencies of all species are very steep inbound ($-$16 to $-$20), but much less steep outbound ($-$4 to $-$12). The outbound {\rh}-dependencies are still very steep compared to water vaporization models (other comets usually have one side steep but the other fairly shallow due to simple seasonal effects). Our model (using the pole solution and matching the observed jet morphology) can match the timing of the apparition lightcurve ($Q$s versus time) and steepness with a cosine-squared dependence on the Sun's zenith distance. This implies a significantly stronger dependence on the angle of insolation than the canonical cosine dependence, and may give insight into the thermal properties of the near-surface region of the source.



Tempel 2 has ``typical'' abundances, consistent with the findings of previous authors. It is relatively dust poor, having a dust-to-gas ratio matching the database averages \citep{ahearn95} for all comets and for those having similar perihelion distances. The dust color is dull, reddening only 7\% per 1000 \AA, and does not exhibit a trend with aperture size. 


In spite of two seasons of intensive, multi-faceted observing campaigns which have yielded numerous discoveries, some aspects of Tempel 2 remain mysteries that will only be resolved by looking up-close with a spacecraft. Tempel 2 is a frequent spacecraft target candidate (cf. \citealt{osip92}) and, should such a mission occur, these data will provide an extremely important baseline to determine the extent to which the comet is physically evolving over time and to what degree the observed oddities are explained by the detailed structure of the nucleus. In the meantime, remote observations can continue to monitor the spin-down of the nucleus's rotation, the time of on-set of significant activity, and the possible precession of the pole. With enough patience, Tempel 2's mysteries will continue to unravel.



\section*{ACKNOWLEDGEMENTS}
We thank Robert Millis, Peter Birch, and Michael A'Hearn for help with observations in 1983 and 1988, and Christopher Henry, Lori Levy, Kevin Walsh, and Wendy Williams for help with imaging in 1999. We are appreciative of Allison Bair's aid in generating some of the tables and for comparison with the results from our forthcoming database paper. We acknowledge the allocation of telescope time at Perth Observatory and the University of Hawaii. M.M.K. is grateful for office space provided by the University of Maryland Department of Astronomy and Johns Hopkins University Applied Physics Laboratory while working on this project. This work has been supported by numerous NASA grants over the years, most recently NNX09AB51G and NNX11AD95G.



\label{lastpage}

\end{singlespace}

\clearpage

\renewcommand{\baselinestretch}{0.9}
\renewcommand{\arraystretch}{1.0}

\begin{deluxetable}{lccccccccc}  
\tabletypesize{\scriptsize}
\tablecolumns{10}
\tablewidth{0pt} 
\setlength{\tabcolsep}{0.03in}
\tablecaption{Photometry observing circumstances and fluorescence efficiencies for Comet 10P/Tempel 2.\,\tablenotemark{a}}
\tablehead{   
  \multicolumn{1}{l}{UT Date}&
  \colhead{$\Delta$$T$\,\tablenotemark{b}}&
  \colhead{$r_\mathrm{H}$}&
  \colhead{$\Delta$}&
  \colhead{$\theta$}&
  \colhead{Phase Adj.}&
  \colhead{$\dot{r}_\mathrm{H}$}&
  \multicolumn{3}{c}{log $L/N$\phantom{99}(erg s$^{-1}$ molecule$^{-1}$)}\\
  \cmidrule(){8-10}
  \colhead{}&
  \colhead{(day)}&
  \colhead{(AU)}&
  \colhead{(AU)}&
  \colhead{($^\circ$)}&
  \colhead{log$A$(0\deg)$f\rho$\tablenotemark{c}\phantom{9}}&
  \colhead{(km s$^{-1}$)}&
  \colhead{\phantom{000}OH\phantom{000}}&
  \colhead{\phantom{000}NH\phantom{000}}&
  \colhead{\phantom{000}CN\phantom{000}}
}
\startdata

1983 Apr 16.48&\phantom{0}$-$46.4&1.467&1.496&\phantom{0}39.6&+0.44&\phantom{0}$-$6.1&$-$15.063&$-$13.423&$-$12.762\\
1983 May 17.45&\phantom{0}$-$15.4&1.391&1.300&\phantom{0}44.0&+0.45&\phantom{0}$-$2.2&$-$15.026&$-$13.480&$-$12.807\\
1983 Jun 9.83&\phantom{0}\phantom{0}+8.0&1.384&1.208&\phantom{0}45.5&+0.46&\phantom{0}+1.2&$-$15.081&$-$13.444&$-$12.821\\
\\
1988 Jun 10.41&\phantom{0}$-$98.5&1.719&0.787&\phantom{0}20.0&+0.29&\phantom{0}$-$9.9&$-$15.086&$-$13.565&$-$12.939\\
1988 Jun 11.38&\phantom{0}$-$97.5&1.713&0.785&\phantom{0}20.5&+0.30&\phantom{0}$-$9.9&$-$15.083&$-$13.562&$-$12.936\\
\\
1999 May 8.30&$-$123.6&1.906&1.013&\phantom{0}19.5&+0.28&\phantom{0}$-$9.7&$-$15.194&$-$13.642&$-$13.033\\
\enddata
\tablenotetext{a} {All parameters were taken at the midpoint of each night's observations, and all photometry was obtained at Lowell Observatory except for 1983 June 9 which was obtained at Perth Observatory.}
\tablenotetext{b} {Time from perihelion.}
\tablenotetext{c} {Nightly adjustment of log \afrho\ values to 0\deg\ phase angle based on assumed phase function (see the text).}
\label{t:phot_circ}
\end{deluxetable}

\begin{deluxetable}{lcrcccccccccc}  
\tabletypesize{\scriptsize}
\tablecolumns{13}
\tablewidth{0pt} 
\setlength{\tabcolsep}{0.05in}
\tablecaption{Summary of Tempel 2 imaging observations and geometric parameters.\,\tablenotemark{a}}
\tablehead{   
  \multicolumn{1}{l}{UT Date}&
  \colhead{UT}&
  \colhead{$\Delta$$T$}&
  \colhead{\rh}&
  \colhead{$\Delta$}&
  \colhead{$\theta$}&
  \colhead{PA of}&
  \colhead{$\Delta$$t$}&
  \colhead{Ecl. Long.}&
  \colhead{Ecl. Long.}&
  \colhead{$\Delta$$m_1$\,\tablenotemark{e}}&
  \colhead{$\Delta$$m_2$\,\tablenotemark{f}}&
  \colhead{Conditions}\\
  \colhead{}&
  \colhead{}&
  \colhead{(day)}&
  \colhead{(AU)}&
  \colhead{(AU)}&
  \colhead{($^\circ$)}&
  \colhead{Sun ($^\circ$)}&
  \colhead{(hr)\,\tablenotemark{b}}&
  \colhead{Earth ($^\circ$)\,\tablenotemark{c}}&
  \colhead{Sun ($^\circ$)\,\tablenotemark{d}}&
  \colhead{}&
  \colhead{}&
  \colhead{}
}
\startdata
1999 Jun 10&\phantom{0}4:34--11:17&$-$90.1&1.731&0.737&10.5&\phantom{00}7.0&--&--&--&--&--&Photometric\\
1999 Jun 11&\phantom{0}4:30--\phantom{0}9:35&$-$89.1&1.726&0.732&10.4&\phantom{00}3.3&--&--&--&--&--&Photometric\\
1999 Jun 22&\phantom{0}7:29--11:08&$-$78.0&1.675&0.685&12.4&325.5&--&--&--&--&--&Thin clouds\\
1999 Jun 23&\phantom{0}3:57--11:04&$-$77.1&1.671&0.682&12.7&323.2&--&--&--&--&--&Photometric\\
1999 Jul 10&\phantom{0}6:55--\phantom{0}9:11&$-$60.1&1.601&0.654&20.9&293.6&--&--&--&--&--&Clouds\\
1999 Jul 16&\phantom{0}5:22--\phantom{0}8:51&$-$54.1&1.579&0.655&24.0&288.3&--&--&--&--&--&Moderate clouds\\
1999 Aug 5&\phantom{0}4:00--\phantom{0}4:36&$-$34.2&1.522&0.690&32.9&277.6&--&--&--&--&--&Clouds\\
1999 Sep 2&\phantom{0}3:34--\phantom{0}4:56&\phantom{0}$-$6.2&1.483&0.800&39.9&269.0&--&--&--&--&--&Clouds\\
\enddata
\tablenotetext{a} {All parameters were taken at the midpoint of each night's observations, and all images were obtained at Lowell Observatory. Parameters specific to the lightcurve are only listed for nights on which a usable lightcurve was obtained.}
\tablenotetext{b}{{\scriptsize Light travel time.}}  
\tablenotetext{c}{Ecliptic longitude of the Earth as seen from the comet.}
\tablenotetext{d}{Ecliptic longitude of the Sun as seen from the comet.}
\tablenotetext{e}{Magnitude necessary to correct from $m_R$($r$,$\Delta$,$\alpha$) to $m_R$(1,1,0) for the night.}
\tablenotetext{f}{Offset (in magnitudes) necessary to make lightcurves on all nights peak at the same magnitude.}
\label{t:imaging_circ}
\end{deluxetable}

\renewcommand{\baselinestretch}{0.7}
\renewcommand{\arraystretch}{1.0}



\begin{deluxetable}{lccccclccccclccccclccc}
\tabletypesize{\scriptsize}
\tablecolumns{22}
\tablewidth{0pt} 
\setlength{\tabcolsep}{0.03in}
\tablecaption{Table of CCD Photometry 2010--2011}
\tablehead{   
  \multicolumn{1}{l}{Date\tablenotemark{a}}&
  \colhead{UT\tablenotemark{b}}&
  \colhead{$m_R$\tablenotemark{c}}&
  \colhead{$m_R$*\tablenotemark{d}}&
  \colhead{}&
  \colhead{}&
  \colhead{Date\tablenotemark{a}}&
  \colhead{UT\tablenotemark{b}}&
  \colhead{$m_R$\tablenotemark{c}}&
  \colhead{$m_R$*\tablenotemark{d}}&
  \colhead{}&
  \colhead{}&
  \colhead{Date\tablenotemark{a}}&
  \colhead{UT\tablenotemark{b}}&
  \colhead{$m_R$\tablenotemark{c}}&
  \colhead{$m_R$*\tablenotemark{d}}&
  \colhead{}&
  \colhead{}&
  \colhead{Date\tablenotemark{a}}&
  \colhead{UT\tablenotemark{b}}&
  \colhead{$m_R$\tablenotemark{c}}&
  \colhead{$m_R$*\tablenotemark{d}}
}
\startdata


Sep 9&\phantom{1}6.972&13.70&14.58&&&Sep 11&\phantom{1}6.642&13.77&14.61&&&Sep 13&11.400&13.77&14.45&&&Oct 16&\phantom{1}9.465&14.58&14.58\\
Sep 9&\phantom{1}6.988&13.70&14.59&&&Sep 11&\phantom{1}6.657&13.77&14.63&&&Sep 13&11.420&13.77&14.45&&&Oct 16&\phantom{1}9.486&14.58&14.60\\
Sep 9&\phantom{1}7.001&13.71&14.64&&&Sep 11&\phantom{1}6.669&13.78&14.64&&&Sep 13&11.720&13.78&14.47&&&Oct 16&\phantom{1}9.607&14.57&14.57\\
Sep 9&\phantom{1}7.745&13.67&14.41&&&Sep 11&\phantom{1}7.051&13.77&14.60&&&Sep 13&11.740&13.79&14.50&&&Oct 16&\phantom{1}9.628&14.58&14.59\\
Sep 9&\phantom{1}7.766&13.67&14.38&&&Sep 11&\phantom{1}7.063&13.78&14.66&&&Sep 13&11.760&13.78&14.49&&&Oct 16&\phantom{1}9.649&14.57&14.55\\
Sep 9&\phantom{1}7.778&13.67&14.40&&&Sep 11&\phantom{1}7.076&13.77&14.61&&&Oct 16&\phantom{1}3.673&14.65&14.74&&&Oct 16&\phantom{1}9.770&14.57&14.55\\
Sep 9&\phantom{1}8.167&13.67&14.35&&&Sep 11&\phantom{1}7.566&13.78&14.67&&&Oct 16&\phantom{1}3.706&14.64&14.71&&&Oct 16&\phantom{1}9.791&14.56&14.55\\
Sep 9&\phantom{1}8.179&13.66&14.36&&&Sep 11&\phantom{1}7.581&13.78&14.66&&&Oct 16&\phantom{1}3.727&14.67&14.82&&&Oct 16&\phantom{1}9.812&14.56&14.53\\
\enddata
\tablenotetext{a} {UT date of observations.}
\tablenotetext{b} {UT at midpoint of the exposure (uncorrected for light travel time).}
\tablenotetext{c} {Observed R-band magnitude (after applying absolute calibrations, extinction corrections, and comparison star corrections) in a circular aperture with radius 5.9 arcsec.}
\tablenotetext{d} {Coma-removed $m_R$(1,1,0) corrected by $\Delta$$m_2$ (given in Table 1) so that all nights have the same peak magnitude.}
\label{t:photometry}
\end{deluxetable}

\begin{deluxetable}{lccc}  
\tabletypesize{\scriptsize}
\tablecolumns{4}
\tablewidth{0pt} 
\setlength{\tabcolsep}{0.05in}
\tablecaption{Sidereal Rotation Periods}
\tablehead{   
  \multicolumn{1}{l}{UT Date\tablenotemark{a}}&
  \colhead{Synodic}&
  \multicolumn{2}{c}{Sidereal Period (hr)}\\
  \cmidrule(lr){3-4}
  \colhead{}&
  \colhead{Period (hr)}&
  \colhead{Prograde}&
  \colhead{Retrograde}
}
\startdata
1988 Apr 28&8.932 $\pm$ 0.001&8.930&8.934\\
1994 Nov 22&8.939 $\pm$ 0.003&8.938&8.940\\
1999 May 20&8.941 $\pm$ 0.002&8.938&8.944\\
2010 Oct 25&8.950 $\pm$ 0.002&8.947&8.953\\
\hline
\enddata
\tablenotetext{a} {Midpoint of the observations used to determine the synodic period.}
\label{t:sidereal_period}
\end{deluxetable}

\begin{deluxetable}{lccccc}  
\tabletypesize{\scriptsize}
\tablecolumns{6}
\tablewidth{0pt} 
\setlength{\tabcolsep}{0.05in}
\tablecaption{Nightly Position Angles}
\tablehead{   
  \multicolumn{1}{l}{UT Date}&
  \multicolumn{2}{c}{\# Meas.\tablenotemark{a}}&
  \multicolumn{3}{c}{Position Angle of Feature ($^\circ$)}\\
  \cmidrule(lr){2-3}
  \cmidrule(lr){4-6}
  \colhead{}&
  \colhead{CN}&
  \colhead{R}&
  \colhead{CN}&
  \colhead{R}&
  \colhead{Model}
}
\startdata
1999 Jun 11&...&2&...&281 $\pm$ 3&328\\
1999 Jun 22&...&2&...&289 $\pm$ 2&328\\
1999 Jun 23&...&2&...&287 $\pm$ 3&328\\
1999 Jul 10&1&2&325 $\pm$ 8&290 $\pm$ 3&327\\
1999 Jul 16&...&1&...&300 $\pm$ 2&327\\
1999 Aug 5&...&2&...&305 $\pm$ 4&326\\
1999 Oct 2&2&1&331 $\pm$ 4&313 $\pm$ 5&323\\
1999 Oct 3&3&2&328 $\pm$ 2&318 $\pm$ 3&323\\
\enddata
\tablenotetext{a} {Number of images for which the position angle was measured.}
\label{t:jet_pas}
\end{deluxetable}

\renewcommand{\baselinestretch}{0.9}
\renewcommand{\arraystretch}{1.0}

\begin{deluxetable}{lllccccccccccccccc}  
\tabletypesize{\scriptsize}
\tablecolumns{18}
\tablewidth{0pt} 
\setlength{\tabcolsep}{0.03in}
\tablecaption{Photometric fluxes and aperture abundances for Comet 10P/Tempel 2}
\tablehead{   
  \multicolumn{3}{l}{UT Date}&
  \multicolumn{2}{c}{Aperture}&
  \multicolumn{5}{c}{log Emission Band Flux}&
  \multicolumn{3}{c}{log Continuum Flux\tablenotemark{a,b}}&
  \multicolumn{5}{c}{log $M$($\rho$)}\\
  \cmidrule(lr){4-5}
  \colhead{}&
  \colhead{}&
  \colhead{}&
  \colhead{Size}&
  \colhead{log $\rho$}&
  \multicolumn{5}{c}{(erg cm$^{-2}$ s$^{-1}$)}&
  \multicolumn{3}{c}{(erg cm$^{-2}$ s$^{-1}$ \AA$^{-1}$)}&
  \multicolumn{5}{c}{(molecule)}\\
  \cmidrule(lr){6-10}
  \cmidrule(lr){11-13}
  \cmidrule(lr){14-18}
  \colhead{}&
  \colhead{}&
  \colhead{}&
  \colhead{(arcsec)}&
  \colhead{(km)}&
  \colhead{OH}&
  \colhead{NH}&
  \colhead{CN}&
  \colhead{C$_3$}&
  \colhead{C$_2$}&
  \colhead{UV}&
  \colhead{Blue}&
  \colhead{Green}&
  \colhead{OH}&
  \colhead{NH}&
  \colhead{CN}&
  \colhead{C$_3$}&
  \colhead{C$_2$}

}
\startdata

1983&Apr&16.48&\phantom{0}20.2&4.04&...&...&...&...&$-$11.80&...&...&$-$14.45&...&...&...&...&28.67\\
1983&May&17.45&\phantom{0}40.1&4.28&...&...&$-$10.84&...&$-$10.82&$-$14.13&...&$-$13.92&...&...&29.65&...&29.50\\
1983&Jun&\phantom{0}9.82&109.9&4.68&...&...&$-$10.12&$-$10.15&$-$10.09&$-$14.41&...&$-$13.29&...&...&30.31&29.75&30.15\\
1983&Jun&\phantom{0}9.84&109.9&4.68&...&...&$-$10.12&$-$10.18&$-$10.06&$-$13.92&...&$-$13.34&...&...&30.31&29.72&30.19\\
\\
1988&Jun&10.30&\phantom{0}10.1&3.46&...&...&...&...&$-$13.22&...&...&$-$14.66&...&...&...&...&26.84\\
1988&Jun&10.35&\phantom{0}10.1&3.46&...&...&...&...&$-$13.26&...&...&$-$14.73&...&...&...&...&26.79\\
1988&Jun&10.43&\phantom{0}10.1&3.46&...&...&...&...&$-$13.29&...&...&$-$14.73&...&...&...&...&26.77\\
\enddata
\tablenotetext{a} {Continuum filter wavelengths: UV (1983/1988) = 3650 \AA; UV (1999/2010) = 3448 \AA; blue (1999/2010) = 4450 \AA; green (1983/1988) = 4845 \AA; green (1999/2010) = 5260 \AA.}
\tablenotetext{b} {``$und$'' stands for ``undefined'' and means the continuum flux was measured but the 1-$\sigma$ upper limit was less than 0.}
\label{t:phot_flux}
\end{deluxetable}

\begin{deluxetable}{lllcccccccccccc}  
\tabletypesize{\scriptsize}
\tablecolumns{15}
\tablewidth{0pt} 
\setlength{\tabcolsep}{0.04in}
\tablecaption{Photometric production rates for Comet 10P/Tempel 2.}
\tablehead{   
  \multicolumn{3}{l}{UT Date}&
  \colhead{$\Delta$$T$}&
  \colhead{log $r_\mathrm{H}$}&
  \colhead{log $\rho$}&
  \multicolumn{5}{c}{log $Q$\tablenotemark{a}\phantom{00}(molecules s$^{-1}$)}&
  \multicolumn{3}{c}{log $A$($\theta$)$f\rho$\tablenotemark{a,b,c}\phantom{0,00}(cm)}&
  \colhead{log $Q$}\\
  \cmidrule(lr){7-11}
  \cmidrule(lr){12-14}
  \cmidrule(){15-15}
  \colhead{}&
  \colhead{}&
  \colhead{}&
  \colhead{(day)}&
  \colhead{(AU)}&
  \colhead{(km)}&
  \colhead{OH}&
  \colhead{NH}&
  \colhead{CN}&
  \colhead{C$_3$}&
  \colhead{C$_2$}&
  \colhead{UV}&
  \colhead{Blue}&
  \colhead{Green}&
  \colhead{H$_2$O}

}
\startdata

1983&Apr&16.48&\phantom{0}$-$46.4&0.166&4.04&...{\tiny\phantom{.00}}&...{\tiny\phantom{.00}}&...{\tiny\phantom{.00}}&...{\tiny\phantom{.00}}&25.17{\tiny\phantom{.}.02}&...{\tiny\phantom{.00}}&...{\tiny\phantom{.00}}&1.87{\tiny\phantom{.}.05}&...\\
1983&May&17.45&\phantom{0}$-$15.4&0.143&4.28&...{\tiny\phantom{.00}}&...{\tiny\phantom{.00}}&25.55{\tiny\phantom{.}.01}&...{\tiny\phantom{.00}}&25.59{\tiny\phantom{.}.00}&1.99{\tiny\phantom{.}.07}&...{\tiny\phantom{.00}}&2.00{\tiny\phantom{.}.02}&...\\
1983&Jun&\phantom{0}9.82&\phantom{0}\phantom{0}+7.9&0.141&4.68&...{\tiny\phantom{.00}}&...{\tiny\phantom{.00}}&25.66{\tiny\phantom{.}.00}&25.21{\tiny\phantom{.}.00}&25.69{\tiny\phantom{.}.00}&1.26{\tiny\phantom{.}.04}&...{\tiny\phantom{.00}}&2.15{\tiny\phantom{.}.01}&...\\
1983&Jun&\phantom{0}9.84&\phantom{0}\phantom{0}+8.0&0.141&4.68&...{\tiny\phantom{.00}}&...{\tiny\phantom{.00}}&25.66{\tiny\phantom{.}.00}&25.18{\tiny\phantom{.}.00}&25.72{\tiny\phantom{.}.00}&1.75{\tiny\phantom{.}.03}&...{\tiny\phantom{.00}}&2.10{\tiny\phantom{.}.01}&...\\
\\
1988&Jun&10.30&\phantom{0}$-$98.6&0.235&3.46&...{\tiny\phantom{.00}}&...{\tiny\phantom{.00}}&...{\tiny\phantom{.00}}&...{\tiny\phantom{.00}}&24.41{\tiny\phantom{.}.10}&...{\tiny\phantom{.00}}&...{\tiny\phantom{.00}}&1.81{\tiny\phantom{.}.01}&...\\
1988&Jun&10.35&\phantom{0}$-$98.6&0.235&3.46&...{\tiny\phantom{.00}}&...{\tiny\phantom{.00}}&...{\tiny\phantom{.00}}&...{\tiny\phantom{.00}}&24.37{\tiny\phantom{.}.11}&...{\tiny\phantom{.00}}&...{\tiny\phantom{.00}}&1.75{\tiny\phantom{.}.02}&...\\
1988&Jun&10.43&\phantom{0}$-$98.5&0.235&3.46&...{\tiny\phantom{.00}}&...{\tiny\phantom{.00}}&...{\tiny\phantom{.00}}&...{\tiny\phantom{.00}}&24.34{\tiny\phantom{.}.12}&...{\tiny\phantom{.00}}&...{\tiny\phantom{.00}}&1.75{\tiny\phantom{.}.02}&...\\
\enddata
\tablenotetext{a} {Production rates, followed by the upper, i.e., the positive uncertainty. The ``$+$'' and ``$-$'' uncertainties are equal as percentages, but unequal in log-space; the ``$-$'' values can be computed.}
\tablenotetext{b} {Continuum filter wavelengths: UV (1983/1988) = 3650 \AA; UV (1999/2010) = 3448 \AA; blue (1999/2010) = 4450 \AA; green (1983/1988) = 4845 \AA; green (1999/2010) = 5260 \AA.}
\tablenotetext{c} {``$und$'' stands for ``undefined'' and means the continuum flux was measured but the 1-$\sigma$ upper limit was less than 0.}
\label{t:phot_rates}
\end{deluxetable}

\begin{deluxetable}{lccccc}  
\tabletypesize{\scriptsize}
\tablecolumns{6}
\tablewidth{0pt} 
\setlength{\tabcolsep}{0.03in}
\tablecaption{Abundance ratios for 10P/Tempel 2}

\tablehead{   
  \colhead{Species}&
  \multicolumn{2}{c}{$r_\mathrm{H}$-dependence\tablenotemark{a}}&
  \multicolumn{3}{c}{log Production Rate}\rule{0cm}{8pt}\\
  \colhead{}&
  \colhead{}&
  \colhead{}&
  \multicolumn{3}{c}{Ratios (X/OH)\tablenotemark{b}}\\
  \cmidrule(lr){2-3}
  \cmidrule(lr){4-6}
  \colhead{}&
  \colhead{Pre-Peri}&
  \colhead{Post-Peri}&
  \colhead{Mean}&
  \colhead{$\sigma$$_\mathrm{mean}$}&
  \colhead{$\sigma$$_\mathrm{data}$}
}
\startdata
OH&$-$16.6$\pm$2.2&\phantom{0}$-$8.6$\pm$0.3&--&--&--\\
NH&$-$20.3$\pm$1.4&$-$13.1$\pm$0.6&\phantom{0}$-$2.32&+0.03&+0.18\\
CN&$-$19.5$\pm$1.1&\phantom{0}$-$3.2$\pm$0.2&\phantom{0}$-$2.44&+0.07&+0.38\\
C$_3$&$-$16.9$\pm$1.6&\phantom{0}$-$4.1$\pm$0.4&\phantom{0}$-$2.92&+0.04&+0.25\\
C$_2$&$-$18.5$\pm$1.1&\phantom{0}$-$4.8$\pm$0.3&\phantom{0}$-$2.34&+0.07&+0.41\\
\\
UC&\phantom{0}$-$1.2$\pm$6.4&\phantom{0}$-$1.9$\pm$0.8&$-$25.54&+0.14&+0.57\\
BC&$-$12.5$\pm$2.7&\phantom{0}$-$2.6$\pm$0.6&$-$25.51&+0.11&+0.46\\
\enddata
\tablenotetext{a} {These \rh-dependencies are based only on the 1999 and 2010 apparitions but exclude the pre-turn on data.}
\tablenotetext{b} {For the dust continuum, the ratio of $Af$$\rho$ to $Q$(OH) has units of cm s mol$^{-1}$. $\sigma$$_\mathrm{data}$ is the standard deviation and measures the scatter of the data values around the sample mean, while $\sigma$$_\mathrm{mean}$ is the standard deviation of the sampling distribution of the mean, or the standard deviation divided by the square root of the number of cases. Only upper error bars are given; the lower error bars can be derived if desired. The dust-to-gas production rates also exclude the pre-turn on data to avoid the contributions from the nucleus.}
\tablenotetext{c} {$A(\theta)f$$\rho$ adjusted to $\theta$ = 0$^\circ$.}
\label{t:abundance_ratios}
\label{lasttable}
\end{deluxetable}

\clearpage

\renewcommand{\baselinestretch}{0.8}
\renewcommand{\arraystretch}{1.0}

\begin{figure}
  \figurenum{4}
  \caption[Example of unwrapping]{Example of unwrapping procedure to measure the position angle (PA) of the feature. The top panel shows a CN image from 2010 September 9 which is centered on the comet and has been enhanced by subtraction of an azimuthal median profile. The image is $\sim$570,000 km on a side. White represents the brightest areas and dark blue/black represents the darkest areas. The faint ring-like structures are artifacts of the enhancement process. 
The middle panel shows the inner portion of the same image after having been ``unwrapped'', rebinned radially and azimuthally, and smoothed. The left edge is PA = 0$^\circ$ and the right edge is PA = 360$^\circ$. The $y$-axis is the distance from the nucleus, $\rho$, with 0 at the bottom and $\sim$71,000 km at the top. The color scale is the same as in the top panel. 
The bottom panel shows line profiles of the unwrapped image that have been rebinned radially by an additional factor of four and are approximately centered on the feature. The line profiles have a different color scale than the images. The vertical scale is brightness; the profiles have a different vertical scale from one another but the color scale is the same on all profiles, with red/orange representing the brightest areas and blue/purple representing the faintest areas. Note that since the feature spans 0$^\circ$/360$^\circ$, PA values greater than 360$^\circ$ are really PA $-$ 360$^\circ$. The first two profiles of the unwrapped image are generally ignored due to effects introduced by the centroiding and enhancement, and in fact the first profile is not plotted here. The profiles are in steps of $\rho$ $\sim$ 7100 km. The CN feature is measured to have a mean PA of 29$^\circ$ $\pm$ 2$^\circ$.}
\end{figure}

\end{document}